\newcommand\SEKImasterusepackages{\usepackage[T1]{fontenc}}
\newcommand\SEKIusersusepackages
{
\usepackage{named}
\bibliographystyle{cpsnamedwithapalikesorting}
}

\newcommand\SEKIREVIEWSTATUS{2}
\newcommand\SEKISUBEDITOR
{{\sc Erica Melis} 
\\FR Informatik, Universit\"at des Saarlandes, D--66123 Saarbr\"ucken, Germany
\\E-mail: {\tt melis@activemath.org}
\\WWW: \url{http://www.activemath.org/~melis}
}

\newcommand\SEKIREVIEWER
{{\sc Erica Melis} 
\\FR Informatik, Universit\"at des Saarlandes, D--66123 Saarbr\"ucken, Germany
\\E-mail: {\tt melis@activemath.org}
\\WWW: \url{http://www.activemath.org/~melis}
}

\newcommand\SEKIREFEREEPROCESS
{\ifcologjournalname\ {\bf 3(4)}, 2016}
\newcommand\SEKIDOCUMENTCLASS{0}
\newcommand\SEKIYEAR{2015}
\newcommand\SEKINUMBEROFYEAR{02}
\newcommand\SEKITYPE{0}
\newcommand\SEKIPUBLISHEDTYPE{0}
\title{The Explicit Definition of Quantifiers via \hilbertsepsilon\
is Confluent and Terminating}\author{\wirthname\\\Institute\\\emailcp}
\date{\small
  Received Nov\,17, 2015.\\
  Accepted after Revision, \Feb\,15, 2016}
\input seki-deckblatt-5
\def\citep{\cite}
\def\citet#1{\citeauthor{#1} \shortcite{#1}}
\newcommand\startcite{{\raise.2ex\hbox{[}}}
\newcommand\stopcite {\raise.2ex\hbox{]}}
\newcommand\citehelper[1]{\startcite #1\stopcite}
\newcommand\makeaciteoftwo[2]
{\citehelper{\citeauthor{#1}, \citeyear{#1}; \citeyear{#2}}}

\input headerhot
\usepackage{amssymb}

\usepackage[T1]{fontenc}

\mathchardef\Gammaoffont="7000
\mathchardef\Gamma="0100
\mathchardef\Deltaoffont="7001
\mathchardef\Delta="0101
\mathchardef\Thetaoffont="7002
\mathchardef\Theta="0102
\mathchardef\Lambdaoffont="7003
\mathchardef\Lambda="0103
\mathchardef\Xioffont="7004
\mathchardef\Xi="0104
\mathchardef\Pioffont="7005
\mathchardef\Pi="0105
\mathchardef\Sigmaoffont="7006
\mathchardef\Sigma="0106
\mathchardef\Upsilonoffont="7007
\mathchardef\Upsilon="0107
\mathchardef\Phioffont="7008
\mathchardef\Phi="0108
\mathchardef\Psioffont="7009
\mathchardef\Psi="0109
\mathchardef\Omegaoffont="700A
\mathchardef\Omega="010A
\mathchardef\itype="017B

\catcode`\@=11

\gdef\allowhyphens{\penalty\@M \hskip\z@skip}

\gdef\set@low@box#1{\setbox\tw@\hbox{,}\setbox\z@\hbox{#1}\dimen\z@\ht\z@
     \advance\dimen\z@ -\ht\tw@
     \setbox\z@\hbox{\lower\dimen\z@ \box\z@}\ht\z@\ht\tw@ \dp\z@\dp\tw@ }
\gdef\set@low@boxsingle#1{\setbox\tw@\hbox{\rm,}\setbox\z@\hbox{#1}\dimen\z@\ht\z@
     \advance\dimen\z@ -\ht\tw@
     \setbox\z@\hbox{\lower\dimen\z@ \box\z@}\ht\z@\ht\tw@ \dp\z@\dp\tw@ }
\gdef\@glqq{%
\ifhmode\edef\@SF{\spacefactor\the\spacefactor}%
\else\let\@SF\empty
\fi
\CheckFamily\font\fraknomath\ifSameFamily ``\relax
\else\CheckFamily\font\swab\ifSameFamily ``\relax
\else\leavevmode\set@low@box{''}\box\z@\kern-.04em\allowhyphens\@SF\relax
\fi\fi}
\gdef\glqq{\protect\@glqq\kern+.07em}
\gdef\@grqq{%
\ifhmode\edef\@SF{\spacefactor\the\spacefactor}%
\else\let\@SF\empty 
\fi 
\CheckFamily\font\fraknomath\ifSameFamily ''\relax
\else\CheckFamily\font\swab\ifSameFamily ''\relax
\else\kern+.07em``\kern.07em\@SF\relax
\fi\fi}
\gdef\grqq{\protect\@grqq}
\gdef\@glq{{\ifhmode \edef\@SF{\spacefactor\the\spacefactor}\else
     \let\@SF\empty \fi \leavevmode
     \set@low@boxsingle{'\/}\box\z@\kern-.04em\allowhyphens\@SF\relax}}
\gdef\glq{\protect\@glq\kern+.07em}
\gdef\@grq{\ifhmode \edef\@SF{\spacefactor\the\spacefactor}\else
     \let\@SF\empty \fi \kern-.0125em`\kern.07em\@SF\relax}
\gdef\grq{\protect\@grq}

\catcode`\@=12

\newcommand\closequotecommanospace{''\nolinebreak\hskip-0.23em,}

\newcommand\closequotecommasmallextraspace{\closequotecommanospace\ \,\,}
\newcommand\closequotecommaextraspace{\closequotecommanospace\   \ \,}

\newcommand\closequotefullstopnospace
                                 {''\nolinebreak\hskip-0.20em\@.}

\makeatletter
\if \@ptsize 0
   \newfont{\scriptscriptscriptgoth}{ygoth scaled 760}
   \newfont{\scriptscriptgoth}{ygoth scaled 833}
   \newfont{\scriptgoth}{ygoth scaled 912}
   \newfont{\gothnomath}{ygoth}
   \newfont{\Goth}{ygoth scaled \magstephalf}
   \newfont{\GOth}{ygoth scaled \magstep1}
   \newfont{\GOTh}{ygoth scaled \magstep2}
   \newfont{\GOTH}{ygoth scaled \magstep3}

   \newfont{\scriptscriptscriptswab}{yswab scaled 760}
   \newfont{\scriptscriptswab}{yswab scaled 833}
   \newfont{\scriptswab}{yswab scaled 912}
   \newfont{\swab}{yswab}
   \newfont{\Swab}{yswab scaled \magstephalf}
   \newfont{\SWab}{yswab scaled \magstep1}
   \newfont{\SWAb}{yswab scaled \magstep2}
   \newfont{\SWAB}{yswab scaled \magstep3}

   \newfont{\scriptscriptscriptfrak}{yfrak scaled 760}
   \newfont{\scriptscriptfrak}{yfrak scaled 833}
   \newfont{\scriptfrak}{yfrak scaled 912}
   \newfont{\fraknomath}{yfrak}
   \newfont{\Frak}{yfrak scaled \magstephalf}
   \newfont{\FRak}{yfrak scaled \magstep1}
   \newfont{\FRAk}{yfrak scaled \magstep2}
   \newfont{\FRAK}{yfrak scaled \magstep3}

   \newfont{\init}{yinit}
   \newfont{\Init}{yinit scaled \magstephalf}
   \newfont{\INit}{yinit scaled \magstep1}
   \newfont{\INIt}{yinit scaled \magstep2}
   \newfont{\INIT}{yinit scaled \magstep3}
\fi
\if \@ptsize 1
   \newfont{\scriptscriptscriptgoth}{ygoth scaled 833}
   \newfont{\scriptscriptgoth}{ygoth scaled 912}
   \newfont{\scriptgoth}{ygoth}
   \newfont{\gothnomath}{ygoth scaled \magstephalf}
   \newfont{\Goth}{ygoth scaled \magstep1}
   \newfont{\GOth}{ygoth scaled \magstep2}
   \newfont{\GOTh}{ygoth scaled \magstep3}
   \newfont{\GOTH}{ygoth scaled \magstep4}

   \newfont{\scriptscriptscriptswab}{yswab scaled 833}
   \newfont{\scriptscriptswab}{yswab scaled 912}
   \newfont{\scriptswab}{yswab}
   \newfont{\swab}{yswab scaled \magstephalf}
   \newfont{\Swab}{yswab scaled \magstep1}
   \newfont{\SWab}{yswab scaled \magstep2}
   \newfont{\SWAb}{yswab scaled \magstep3}
   \newfont{\SWAB}{yswab scaled \magstep4}

   \newfont{\scriptscriptscriptfrak}{yfrak scaled 833}
   \newfont{\scriptscriptfrak}{yfrak scaled 912}
   \newfont{\scriptfrak}{yfrak}
   \newfont{\fraknomath}{yfrak scaled \magstephalf}
   \newfont{\Frak}{yfrak scaled \magstep1}
   \newfont{\FRak}{yfrak scaled \magstep2}
   \newfont{\FRAk}{yfrak scaled \magstep3}
   \newfont{\FRAK}{yfrak scaled \magstep4}

   \newfont{\init}{yinit scaled \magstephalf}
   \newfont{\Init}{yinit scaled \magstep1}
   \newfont{\INit}{yinit scaled \magstep2}
   \newfont{\INIt}{yinit scaled \magstep3}
   \newfont{\INIT}{yinit scaled \magstep4}
\fi
\if \@ptsize 2
   \newfont{\scriptscriptscriptgoth}{ygoth scaled 912}
   \newfont{\scriptscriptgoth}{ygoth}
   \newfont{\scriptgoth}{ygoth scaled \magstephalf}
   \newfont{\gothnomath}{ygoth scaled \magstep1}
   \newfont{\Goth}{ygoth scaled \magstep2}
   \newfont{\GOth}{ygoth scaled \magstep3}
   \newfont{\GOTh}{ygoth scaled \magstep4}
   \newfont{\GOTH}{ygoth scaled \magstep5}

   \newfont{\scriptscriptscriptswab}{yswab scaled 912}
   \newfont{\scriptscriptswab}{yswab}
   \newfont{\scriptswab}{yswab scaled \magstephalf}
   \newfont{\swab}{yswab scaled \magstep1}
   \newfont{\Swab}{yswab scaled \magstep2}
   \newfont{\SWab}{yswab scaled \magstep3}
   \newfont{\SWAb}{yswab scaled \magstep4}
   \newfont{\SWAB}{yswab scaled \magstep5}

   \newfont{\scriptscriptscriptfrak}{yfrak scaled 833}
   \newfont{\scriptscriptfrak}{yfrak}
   \newfont{\scriptfrak}{yfrak scaled \magstephalf}
   \newfont{\fraknomath}{yfrak scaled \magstep1}
   \newfont{\Frak}{yfrak scaled \magstep2}
   \newfont{\FRak}{yfrak scaled \magstep3}
   \newfont{\FRAk}{yfrak scaled \magstep4}
   \newfont{\FRAK}{yfrak scaled \magstep5}

   \newfont{\init}{yinit scaled \magstep1}
   \newfont{\Init}{yinit scaled \magstep2}
   \newfont{\INit}{yinit scaled \magstep3}
   \newfont{\INIt}{yinit scaled \magstep4}
   \newfont{\INIT}{yinit scaled \magstep5}
\fi

\newcommand{\mscriptscriptfrak}      [1]{\mbox{\scriptscriptscriptfrak#1}}
\newcommand{\mscriptfrak}            [1]{\mbox{\scriptscriptscriptfrak#1}}

\newcommand{\mfrak}[1]{\mbox{\fraknomath#1}}

\makeatother
\newif\ifSameFamily
\def\CheckFamily#1#2{\GetFamilyName{#1}\ArgOne
        \GetFamilyName{#2}\ArgTwo
        \ifx\ArgOne\ArgTwo\SameFamilytrue\else\SameFamilyfalse\fi}
\def\GetFamilyName#1{\edef\Tempa{#1}\def\Tempb{#1}\ifx\Tempa\Tempb
        \edef\Tempa{\fontname#1}\fi
        \edef\Tempa{\Tempa\space}%
        \expandafter\iGetFamilyName\Tempa\\}
\def\iGetFamilyName#1 #2\\#3{\def#3{#1}}
\def\DefFontName#1#2{{\escapechar-1\expandafter\expandafter\expandafter
        \iDefFontName\expandafter{\csname#2\endcsname}%
        \xdef#1{\expandafter\string\Tempa}}}
\def\iDefFontName{\def\Tempa}

\newcommand\unprotectedae
{\CheckFamily\font\fraknomath\ifSameFamily *a\else
 \CheckFamily\font\swab\ifSameFamily\char'212\else\"a\fi\fi}
\newcommand\unprotectedoe
{\CheckFamily\font\fraknomath\ifSameFamily 
*o\else\CheckFamily\font\swab\ifSameFamily\char'232\else\"o\fi\fi}
\newcommand\unprotectedue
{\CheckFamily\font\fraknomath\ifSameFamily 
*u\else\CheckFamily\font\swab\ifSameFamily\char'237\else\"u\fi\fi}

\DefFontName\eccclarge{eccc1200}
\DefFontName\eccc{eccc1000}
\DefFontName\ecccsmall{eccc0900}
\DefFontName\ecccfootnotesize{eccc0800}
\newcommand\unprotectedsz
{\CheckFamily\font\fraknomath\ifSameFamily\char'032\else
 \CheckFamily\font\swab\ifSameFamily\char'032\else
 \CheckFamily\font\eccclarge\ifSameFamily s%\hskip-.15em 
 z\else 
 \CheckFamily\font\eccc\ifSameFamily s%\hskip-.15em 
 z\else
 \ss\fi\fi\fi\fi%\fi\fi
}
\newcommand\unprotectedes
{\CheckFamily\font\fraknomath\ifSameFamily\char'215\else
\CheckFamily\font\swab\ifSameFamily\char'215\else  
s\fi\fi}
\newcommand\unprotectedesi
{\CheckFamily\font\fraknomath\ifSameFamily\char'215\else
\CheckFamily\font\swab\ifSameFamily\char'215\else  
\mbox{s\hskip.04em}\fi\fi
\-\ignorespaces}

\newcommand\unprotectedmyparagraphsymbol
{\CheckFamily\font\fraknomath\ifSameFamily 
\char'244\else\CheckFamily\font\swab\ifSameFamily
\char'244\else\S\fi\fi}

\renewcommand\ae{\protect\unprotectedae}
\renewcommand\oe{\protect\unprotectedoe}
\newcommand\ue  {\protect\unprotectedue}

\newcommand\sz  {\protect\unprotectedsz}
\newcommand\es  {\protect\unprotectedes}

\newcommand\esi {\protect\unprotectedesi}  %wortinterne Version
\newcommand\fti {\discretionary{f-} {t}{f\mbox {\hskip.08em}t}}

\newcommand\myparagraphsymbol{\protect\unprotectedmyparagraphsymbol}

\def\mathfrak#1{%
\mathchoice
{{\mfrak{#1}}}%                 display-style
{{\mfrak{#1}}}%                 text-style
{{\mscriptfrak{#1}}}%           scriptstyle
{{\mscriptscriptfrak{#1}}}%     scriptscriptstyle
}

\newcommand\namefont{}

\usepackage{url}
\newcommand\majorfootroom{\raisebox{-1.9ex}{\rule{0ex}{.5ex}}}

\newcommand\footroom{\raisebox{-1.5ex}{\rule{0ex}{.5ex}}}

\newcommand\smallfootroom{\raisebox{-1.0ex}{\rule{0ex}{3ex}}}

\newcommand\majorheadroom{\rule{0ex}{3.2ex}}
\newcommand\headroom{\rule{0ex}{2.8ex}}
\newcommand\mediumheadroom{\rule{0ex}{2.4ex}}

\newcommand\claus    {Clau\es}

\newcommand\jan      {Jan}

\newcommand\paul     {Paul}
\newcommand\peter    {Peter}

\newcommand\bernaysindex    {\index{Bernays, Paul (1888--1977)}}

\newcommand\bernaysplain    {\mbox{Ber\-nay\es}}
\newcommand\bernays         {{\namefont\bernaysplain}}
\newcommand\bernaysnameplain     {\bernaysindex\paul\       \bernaysplain}
\newcommand\bernaysname     {{\namefont\bernaysnameplain     }}

\newcommand\bernayslifetime {(1888--1977)}

\newcommand\church          {{\namefont Church}}

\newcommand\fermatbirthyear 
{160?}

\newcommand\secondincompletenesstheorem
{second \incompletenesstheorem}
\newcommand\secondIncompletenessTheorem
{Second \IncompletenessTheorem}

\newcommand\firstincompletenesstheorem
{first \incompletenesstheorem}
\newcommand\firstIncompletenessTheorem
{First \IncompletenessTheorem}

\newcommand\incompletenesstheorem{incompleteness theorem}
\newcommand\IncompletenessTheorem{Incompleteness Theorem}

\newcommand\hilbertplain    {Hilbert}
\newcommand\hilbert         {\mbox{\namefont\hilbertplain}}

\newcommand\hilbertsepsilonlongindex{\index{Hilbert!'s epsilon}}

\newcommand\hilbertsepsilon
                     {\hilbertsepsilonlongindex\hilbert'\es\ \nlbmath\varepsilon}
\newcommand\hilbertbernaysplain{\hilbertplain--\bernaysplain}
\newcommand\hilbertbernays  {{\namefont\hilbertbernaysplain}}

\newcommand\klop            {{\namefont Klop}}
\newcommand\klopname        {{\namefont\jan\ Willem \klop}}

\newcommand\newman          {{\namefont Newman}}

\newcommand\newmanlemma     {\index{Newman Lemma}\newman\ Lemma}

\newcommand\axiomofchoice   {\index{choice!Axiom of Choice}Axiom of Choice}

\newcommand\wirthindex                 {\index{Wirth, Claus-Peter (*1963)}}

\newcommand\wirth           {{\namefont Wirth}}
\newcommand\wirthnamenoindex{{\namefont\claus-\peter\ \wirth}}
\newcommand\wirthname       {\wirthindex\wirthnamenoindex}

\newcommand\afortiori{a fortiori}

\newcommand\f    {\mbox{}{f.}}   %extra block blocks ligature to VI\f
\newcommand\Feb  {Feb.}

\newcommand\Math {Math.}

\newcommand\qedhelp[1]{Q.e.d.~({#1})}
\newcommand\getittotheright[1]  
{\hfill\mbox{}\penalty 100\mbox{\ \,}\nolinebreak
\hfill\hfill\hfill\nolinebreak\mbox{#1}\ignorespaces}

\newcommand\Qeddouble[1]{\underline{\underline{\qedhelp{#1}}}}

\newcommand\Qedbf    [1]{\mbox{\bf\qedhelp{#1}}}

\newcommand\QEDdouble[1]{\getittotheright{\Qeddouble{#1}}}

\newcommand\QEDbf    [1]{\getittotheright{\Qedbf    {#1}}}

\newcommand\theo {Theorem}

\newcommand\WWW  {WWW}

\newcommand\asfollows{\mbox{as follows}}

\newcommand\aswell{as \nolinebreak well}
\newcommand\aswellas{\aswell\ \nolinebreak as}
\newcommand\Cf   {Cf.}
\newcommand\cf   {cf.}

\newcommand\Cfnlb{\Cf\nolinebreak}
\newcommand\cfnlb{\cf\nolinebreak}

\newcommand\CS   {Computer \Sci}

\newcommand\Dept {Dept.}

\newcommand\eg   {e.g.}

\newcommand\firstorder{first-order}

\newcommand\grundlagendermathematikindex{\index{Grundlagen!der Mathematik}}
\newcommand\grundlagendermathematiknoindex{Grund\-lagen der Mathematik}
\newcommand\grundlagendermathematik
                  {\grundlagendermathematikindex\grundlagendermathematiknoindex}

\newcommand\ie   {i.e.}

\newcommand\udiff{\ if\ }

\def\note{Note}

\newcommand\p    {p.}
\newcommand\pp   {pp.}
\newcommand\PP[2]{\pp\,\ignorespaces#1--\ignorespaces#2}
\newcommand\PhD  {PhD}%{Ph.D.}
\newcommand\PhDthesis{\PhD\ thesis}

\newcommand\secondorder{second-order}

\newcommand\sect {\myparagraphsymbol} 
\newcommand\Sci  {Sci.}

\newcommand\singulary{singulary}

\newcommand\wellfounded{well-founded}

\newcommand\wellfoundedness{well-founded\-ness}

\newcommand\WellFoundedness{Well-Founded\-ness}

\newcommand\wrog {w.l.o.g.} % My favorite misspelling. Moreover \wlog is
\newcommand\wrt  {w.r.t.}

\newcommand\Vgl{Vgl.}

\newcommand\thewordand{and}
\newcommand\litspageref[1]{Page\,#1}

\newcommand\litnoteref[1]{\note\,#1}

\newcommand\itemname{item}
\newcommand\Itemname{Item}
\newcommand\litItemref[1]{\Itemname\,#1}
\newcommand\lititemref[1]{\itemname\,#1}

\newcommand\littheoref[1]{\theo\,#1}
\newcommand\litsectref[1]{\sect\,#1} %previously called \sectimm

\newcommand\lititemfromtoref[2]{\itemname s #1 \nolinebreak to \nolinebreak #2}

\newcommand\Examplename{Ex\-am\-ple}
\newcommand\litexamref[1]{\Examplename\,#1}

\newcommand\litlemmref[1]{Lem\-ma\,#1}

\newcommand\litcororef[1]{Corol\-lary\,#1}

\newcommand\littheorefs[2]
{Theorems        \nolinebreak #1 \thewordand\ \nolinebreak #2}

\newcommand\cororef[1]{\litcororef{\ref{#1}}}
\newcommand\lemmref[1]{\litlemmref{\ref{#1}}}

\newcommand\theoref[1]{\littheoref{\ref{#1}}}

\newcommand\sectref[1]{\litsectref{\ref{#1}}}

\newcommand\theorefs[2]{\littheorefs{\ref{#1}}{\ref{#2}}}

\newcommand\nthpositioner[2]
{#1\raisebox{0.52ex}{\scriptsize\hspace{0.07em}#2}}
\newcommand\nth[1]{\nthtinypositioner{#1}{\nthstring{#1}}}
\newcommand\nthtinypositioner[2]{#1\raisebox{0.52ex}{\tiny\hspace{0.07em}#2}}

\newcommand\mthpositioner[2]
{\math{#1}\raisebox{0.52ex}{\scriptsize\hspace{0.07em}#2}}
\newcommand\modulointocountzero[2]
{\count1=#1
\count2=#2
\count0=\count1
\divide  \count0 by \count2
\multiply\count0 by-\count2
\advance \count0 by \count1}
\newcommand\absolutevalueintocountzero[1]
{\count0=#1
\ifnum\count0<0\multiply\count0 by -1\fi}
\newcommand\nthstring[1]
{\def\myargone{#1}\ifcat a\myargone th\else\nthstringnochar{#1}\fi}
\newcommand\nthstringnochar[1]
{\absolutevalueintocountzero{#1}%
\modulointocountzero{\count0}{100} %we need the blank here!
\ifnum\count0>9\ifnum\count0<20 th\else\stupidnthstring\fi
                                  \else\stupidnthstring\fi}
\newcommand\stupidnthstring
{\modulointocountzero{\count0}{10}
\ifnum\count0=1 \hskip-0.2em st\else
\ifnum\count0=2 nd\else
\ifnum\count0=3 rd\else 
                th\fi\fi\fi}

\newcommand\writeascents
[1]{\count4=#1
\ifnum\count4<0 
-\multiply\count4 by -1\fi
\modulointocountzero{\count4}{10}
\divide\count4 by 10
\count3=\the\count0
\modulointocountzero{\count4}{10}
\divide\count4 by 10
\the\count4
.\the\count0
\the\count3
}

\newcommand\frenchnthstring[1]
{\def\myargone{#1}\ifcat a\myargone th\else\frenchnthstringnochar{#1}\fi}
\newcommand\frenchnthstringnochar[1]
{\absolutevalueintocountzero{#1}%
\modulointocountzero{\count0}{100} %we need the blank here!
\ifnum\count0>9\ifnum\count0<20 th\else\frenchstupidnthstring\fi
                                  \else\frenchstupidnthstring\fi}
\newcommand\frenchstupidnthstring
{\modulointocountzero{\count0}{10}
\ifnum\count0=1 \hskip-0.2em re\else
\ifnum\count0=2 me\else
\ifnum\count0=3 rd\else 
                th\fi\fi\fi}

\newcommand\CLAM      {{\rm CL\kern-.36em\raise.39ex\hbox{\sc a}\kern-.15emM}}

\newcommand\TEXMACS   {{\sc T\kern-.1667em\lower.5ex\hbox{E}\kern-.125emX\kern-.1em\lower.5ex\hbox{\textsc{m\kern-.05ema\kern-.125emc\kern-.05ems}}}}

\newcommand\Zuerich        {\mbox{Z\ue rich}}

\newcommand\plzETHZ{\mbox{8092}}

\newcommand\ETHshort{ETH}
\newcommand\ETHZshort{\ETHshort\ Zurich}

\def       \emailcp      {{\tt wirth@logic.at}}

\newcommand\Institutedept
{\Dept\ of \Math}
\newcommand\Instituteinst
{\mbox\ETHZshort}
\newcommand\Instituteplac
{\plzETHZ\,\Zuerich}
\newcommand\Institutecoun
{Switzerland}
\newcommand\Institutestre
{R\ae mistr.\,101}
\newcommand\Institute
{\Institutedept, \Instituteinst, \Institutestre, \Instituteplac, \Institutecoun}
\newcommand\academicpress{Academic Press (\elsevier)}

\newcommand\elsevier{Elsevier}

\newcommand\newspaperreference[5]
{\def\nameofjournalpress{#2}#1, #4 #5, #3\if?\nameofjournalpress
\else, #2\fi}

\newcommand\dateinjournal[1]{}

\newcommand\journalreference[6]
{\def\nameofjournalpress{#2}#1\nolinebreak\hskip.2em%
\dateinjournal{(#3) }{\mbox{\bf #4}}, \PP{#5}{#6}\if?\nameofjournalpress
\else, #2\fi}

\newcommand\journalreferenceprintyear[6]
{\def\nameofjournalpress{#2}#1 
#4:#5--#6, #3%
\if?\nameofjournalpress
\else, #2\fi}

\newcommand\journalreferenceprintyearaspartofnumber[6]
{\def\nameofjournalpress{#2}#1 
{#4/#3}, \PP{#5}{#6}\if?\nameofjournalpress
\else, #2\fi}

\newcommand\ifcologjournalname{IfCoLog Journal of Logics and their Applications}

\newcommand\jscname
{J. Symbolic Computation}

\newcommand\jscprintyear
{\journalreferenceprintyear{\jscname}\academicpress}

\newcommand\tcsname{Theoretical \CS}
\newcommand\tcsjournal
{\journalreference\tcsname\elsevier}
\newcommand\tcsjournalprintyear
{\journalreferenceprintyear\tcsname\elsevier}

\urldef\wirthkuhnurl\url
{http://wirth.bplaced.net/SEKI/welcome.html#SWP-2007-01}

\urldef\urlsrninetythreedashzerofive\url
{http://wirth.bplaced.net/SEKI/welcome.html#SR-93-05}

\urldef\urlsreightyeighttwelve\url
{http://wirth.bplaced.net/SEKI/welcome.html#SR-88-12}
\mathcommand\ident[1]{\mathsf{#1}}
\newcommand\plussymbol  {\ident{+}}
\newcommand\minussymbol {\ident{-}}
\newcommand\dividesymbol{\ident{/}}
\newcommand\timessymbol {\ident{*}}

\newcommand\set     {\ident{set}}

\newcommand\naturalssymbol{\ident{naturals}}
\newcommand\gensymsymbol{\ident{gensym}}
\mathcommand\mbpsymbol{\ident{m\hspace{-0.055em}b\hspace{-0.045em}p}}

\newcommand\csymbol     {\ident c}
\newcommand\esymbol     {\ident e}
\newcommand\fsymbol     {\ident f}
\newcommand\gsymbol     {\ident g}
\newcommand\hsymbol     {\ident h}
\newcommand\ksymbol     {\ident k}
\newcommand\psymbol     {\ident p}
\newcommand\ssymbol     {\ident s}
\newcommand\Everysymbol {\ident{Every}}
\newcommand\Permsymbol {\ident{Perm}}
\newcommand\RExistssymbol{\ident{Rexists}}
\newcommand\invertsymbol{\ident{invert}}
\newcommand\invsymbol{\ident{inv}}
\newcommand\abssymbol   {\ident{abs}}
\newcommand\cnssymbol   {\ident{cons}}
\mathcommand\cnsindexsymbol[1]{\ident{cons}_{#1}}
\newcommand\carsymbol   {\ident{car}}

\newcommand\cdrsymbol   {\ident{cdr}}
\newcommand\lengthsymbol{\ident{length}}
\newcommand\sizesymbol{\ident{size}}
\newcommand\dlsymbol    {\ident{dl}}
\newcommand\dloncesymbol{\ident{delfirst}}
\newcommand\rcsymbol    {\ident{rc}}
\newcommand\brsymbol    {\ident{br}}
\newcommand\revtailsymbol{\ident{revtail}}
\newcommand\revsymbol{\ident{rev}}
\newcommand\appendsymbol {\ident{append}}
\newcommand\zeropredicatesymbol{\ident{zerop}}
\newcommand\eqsymbol        {\ident{eq}}
\newcommand\ifthensymbol    {\mbox{\ident{If{}Then}}}
\newcommand\ifthenelsesymbol{\mbox{\ident{If{}ThenElse}}}
\mathcommand\eqindexsymbol        [1]{\eqsymbol        _{#1}}
\mathcommand\ifthenindexsymbol    [1]{\ifthensymbol    _{#1}}
\mathcommand\ifthenelseindexsymbol[1]{\ifthenelsesymbol_{#1}}
\newcommand\orsymbol    {\ident{or}}
\newcommand\andsymbol   {\ident{and}}
\newcommand\leqsymbol   {\ident{leq}}
\newcommand\lessymbol   {\ident{less}}
\newcommand\lexlessymbol{\ident{lexless}}
\newcommand\lexlimlessymbol{\ident{lexlimless}}
\newcommand\lexsymbol   {\ident{lex}}
\newcommand\acksymbol   {\ident{ack}}
\newcommand\switchsymbol{\ident{switch}}
\newcommand\swatchsymbol{\ident{swatch}}
\newcommand\diveinssymbol{\ident{div1}}
\newcommand\divzweisymbol{\ident{div2}}
\newcommand\divrestsymbol{\ident{divrest}}
\newcommand\diveinstailsymbol{\ident{div1tail}}
\newcommand\divzweitailsymbol{\ident{div2tail}}
\newcommand\remsymbol{\ident{rem}}
\newcommand\divsymbol{\ident{div}}

\newcommand\turingmachinesymbol{\ident T}
\newcommand\terminatespsymbol  {\ident{terminatesp}}
\newcommand\statesymbol        {\ident{state}}
\newcommand\cmdsymbol          {\ident{cmd}}
\newcommand\nthsymbol          {\ident{nth}}
\newcommand\doublesymbol       {\ident{double}}

\newcommand\ppsymbol           {\ident{p}}
\newcommand\qpsymbol           {\ident{q}}
\newcommand\Epsymbol           {\ident{E}}
\newcommand\Ppsymbol           {\ident{P}}
\newcommand\Qpsymbol           {\ident{Q}}
\newcommand\Marriessymbol      {\ident{Marries}}
\newcommand\Lovessymbol        {\ident{Loves}}
\newcommand\StolenBysymbol     {\ident{StolenBy}}
\newcommand\Humansymbol        {\ident{Human}}
\newcommand\Evensymbol         {\ident{Even}}
\newcommand\Oddsymbol          {\ident{Odd}}
\newcommand\Primesymbol        {\ident{Prime}}
\newcommand\EveryPairsymbol   {\ident{EveryPair}}
\newcommand\Givesymbol         {\ident{Give}}
\newcommand\Fathersymbol       {\ident{Father}}
\newcommand\Elephantpsymbol    {\ident{Elephant}}
\newcommand\Flowerpsymbol    {\ident{Flower}}
\newcommand\Germanpsymbol      {\ident{German}}
\newcommand\Bicyclepsymbol     {\ident{Bicycle}}
\newcommand\Hugepsymbol        {\ident{Huge}}
\newcommand\Animalpsymbol      {\ident{Animal}}
\newcommand\Malepsymbol        {\ident{Male}}
\newcommand\Boypsymbol         {\ident{Boy}}
\newcommand\Girlpsymbol        {\ident{Girl}}
\newcommand\Femalepsymbol      {\ident{Female}}
\newcommand\Roundpsymbol       {\ident{Round}}
\newcommand\Quadrangularpsymbol{\ident{Quadrangular}}
\newcommand\Metpsymbol         {\ident{Met}}
\newcommand\Kissedpsymbol      {\ident{Kissed}}
\newcommand\Bishopsymbol       {\ident{Bishop}}
\newcommand\mindexsymbol[1]{\existsvari w{#1}}

\newcommand\nonnegpsymbol      {\ident{nonnegp}}
\newcommand\wellsymbol         {\ident{well}}
\newcommand\welltailsymbol     {\ident{welltail}}
\newcommand\varsymbol          {\ident{var}}
\newcommand\aritysymbol        {\ident{arity}}

\newcommand\whilesymbol        {\ident{while}}

\newcommand\nullsymbol         {\ident{null}}
\newcommand\hdsymbol           {\ident{hd}}
\newcommand\tlsymbol           {\ident{tl}}
\newcommand\insymbol           {\ident{in}}
\newcommand\applysymbol        {\ident{app}}
\newcommand\termsymbol         {\ident{term}}
\newcommand\russellsymbol      {\ident{russell}}
\newcommand\sqrtindordsymbol[1]{\ident{sqrtio#1}}
\mathcommand\tightim{\longrightarrow}
\mathcommand\im{\ \tightim\ }
\mathcommand\rs{\:\rulesugar\:\:}
\mathcommand\rulesugar{\longleftarrow}

\mathcommand\doublepp[1]      {\doublesymbol   \beginargs{#1}\allargs}
\mathcommand\aritypp[1]      {\aritysymbol   \beginargs{#1}\allargs}
\mathcommand\lengthpp[1]      {\lengthsymbol   \beginargs{#1}\allargs}
\mathcommand\sizepp[1]      {\sizesymbol   \beginargs{#1}\allargs}
\mathcommand\wellpp[1]      {\wellsymbol   \beginargs{#1}\allargs}
\mathcommand\welltailpp[1]      {\welltailsymbol   \beginargs{#1}\allargs}
\mathcommand\varpp[1]      {\varsymbol   \beginargs{#1}\allargs}
\mathcommand\rempp[2]    {\remsymbol\beginargs{#1}\separgs{#2}\allargs}
\mathcommand\divpp[2]    {\divsymbol\beginargs{#1}\separgs{#2}\allargs}
\mathcommand\divrestpp[2]    {\divrestsymbol\beginargs{#1}\separgs{#2}\allargs}
\mathcommand\diveinspp[2]    {\diveinssymbol\beginargs{#1}\separgs{#2}\allargs}
\mathcommand\divzweipp[3]    {\divzweisymbol\beginargs{#1}\separgs{#2}
\separgs{#3}\allargs}
\mathcommand\diveinstailpp[4]    {\diveinstailsymbol\beginargs{#1}\separgs{#2}
\separgs{#3}\separgs{#4}\allargs}
\mathcommand\divzweitailpp[6]    {\divzweitailsymbol\beginargs{#1}\separgs{#2}
\separgs{#3}\separgs{#4}\separgs{#5}\separgs{#6}\allargs}
\mathcommand\mbppp[2]         {\mbpsymbol   \beginargs{#1}\separgs{#2}\allargs}
\mathcommand\revpp[1]     
{\revsymbol\beginargs{#1}\allargs}
\mathcommand\revppi[2]     
{\revsymbol^{#1}\beginargs{#2}\allargs}
\mathcommand\revtailpp[2]     
{\revtailsymbol\beginargs{#1}\separgs{#2}\allargs}
\mathcommand\revtailppi[3]
{\revtailsymbol^{#1}\beginargs{#2}\separgs{#3}\allargs}
\mathcommand\Permpp[2]     
{\Permsymbol\beginargs{#1}\separgs{#2}\allargs}
\mathcommand\Permppi[3]
{\Permsymbol^{#1}\beginargs{#2}\separgs{#3}\allargs}
\mathcommand\appendpp[2]      
{\appendsymbol \beginargs{#1}\separgs{#2}\allargs}
\mathcommand\appendppi[3]      
{\appendsymbol^{#1}\beginargs{#2}\separgs{#3}\allargs}
\mathcommand\Everypp[2]      
{\Everysymbol \beginargs{#1}\separgs{#2}\allargs}
\mathcommand\RExistspp[1]      
{\RExistssymbol \beginargs{#1}\allargs}
\mathcommand\appendlongpp[2]      
{\appendsymbol\left(\begin{array}{@{}l@{}}{#1}\separgs\\{#2}\end{array}\right)}
\mathcommand\cnspp[2]         {\cnssymbol   \beginargs{#1}\separgs{#2}\allargs}
\mathcommand\cnsppi[3]       {\cnssymbol^{#1}\beginargs{#2}\separgs{#3}\allargs}
\mathcommand\cnsindexpp[3]
{\cnsindexsymbol{#1}\beginargs{#2}\separgs{#3}\allargs}
\mathcommand\dlpp[2]          {\dlsymbol    \beginargs{#1}\separgs{#2}\allargs}
\mathcommand\dloncepp[2]      {\dloncesymbol\beginargs{#1}\separgs{#2}\allargs}
\mathcommand\dlonceppi[3]{\dloncesymbol^{#1}\beginargs{#2}\separgs{#3}\allargs}
\mathcommand\rcpp[2]          {\rcsymbol    \beginargs{#1}\separgs{#2}\allargs}
\mathcommand\brpp[2]          {\brsymbol    \beginargs{#1}\separgs{#2}\allargs}
\mathcommand\orpp[2]          {\orsymbol    \beginargs{#1}\separgs{#2}\allargs}
\mathcommand\andpp[2]         {\andsymbol   \beginargs{#1}\separgs{#2}\allargs}
\mathcommand\shortcnspp[2]    {\csymbol     \beginargs{#1}\separgs{#2}\allargs}
\mathcommand\tightshortcnspp[2]
{\csymbol\beginargs{#1}\tightsepargs{#2}\allargs}
\mathcommand\spp[1]           {\ssymbol     \beginargs{#1}\allargs}
\mathcommand\sppiterated[2]   {\ssymbol^{#1}\beginargs{#2}\allargs}
\mathcommand\sqrtindordpp[3]
                       {\sqrtindordsymbol{#1}\beginargs{#2}\separgs{#3}\allargs}
\mathcommand\ppp[1]           {\psymbol     \beginargs{#1}\allargs}
\mathcommand\pppiterated[2]   {\psymbol^{#1}\beginargs{#2}\allargs}
\mathcommand\zeropp           {\ident 0}
\mathcommand\Julietpp         {\ident{Juliet}}
\mathcommand\Romeopp          {\ident{Romeo}}
\mathcommand\Ipp              {\ident I}
\mathcommand\onepp            {\ident1}
\mathcommand\twopp            {\ident2}
\mathcommand\threepp          {\ident3}
\mathcommand\invertpp[1]      {\invertsymbol\beginargs{#1}\allargs}
\mathcommand\invpp[1]         {\invsymbol\beginargs{#1}\allargs}
\mathcommand\abspp[1]         {\abssymbol\beginargs{#1}\allargs}
\mathcommand\naturalspp[1]    {\naturalssymbol\beginargs{#1}\allargs}
\mathcommand\gensympp[1]      {\gensymsymbol\beginargs{#1}\allargs}
\mathcommand\nilpp            {\ident{nil}}
\mathcommand\falsepp          {\ident{false}}
\mathcommand\truepp           {\ident{true}}
\mathcommand\FALSEpp          {\ident{FALSE}}
\mathcommand\TRUEpp           {\ident{TRUE}}
\mathcommand\UNDEFpp          {\ident{UNDEF}}
\mathcommand\weirdppp         {\ident{weirdp}}
\mathcommand\ambigppp         {\ident{ambigp}}
\mathcommand\zeropredicatepp[1]{\zeropredicatesymbol\beginargs{#1}\allargs}
\mathcommand\cppeins       [1]{\csymbol     \beginargs{#1}\allargs}
\mathcommand\cppzwei       [2]{\csymbol\beginargs{#1}\separgs{#2}\allargs}
\mathcommand\eppeins       [1]{\esymbol     \beginargs{#1}\allargs}
\mathcommand\fppeins       [1]{\fsymbol     \beginargs{#1}\allargs}
\mathcommand\fppeinsindex  [2]{\fsymbol_{#1}\beginargs{#2}\allargs}
\mathcommand\fppeinsiterated[2]{\fsymbol^{#1}\beginargs{#2}\allargs}
\mathcommand\gppeins       [1]{\gsymbol     \beginargs{#1}\allargs}
\mathcommand\gppzwei       [2]{\gsymbol     \beginargs{#1}\separgs{#2}\allargs}
\mathcommand\hppeins       [1]{\hsymbol     \beginargs{#1}\allargs}
\mathcommand\kppeins       [1]{\ksymbol     \beginargs{#1}\allargs}
\mathcommand\appzero          {\ident a}
\mathcommand\bppzero          {\ident b}
\mathcommand\cppzero          {\ident c}
\mathcommand\dppzero          {\ident d}
\mathcommand\eppzero          {\ident e}
\mathcommand\eqindexpp[3]{\eqindexsymbol{#1}\beginargs{#2}\separgs{#3}\allargs}
\mathcommand\ifthenindexpp
[3]{\ifthenindexsymbol{#1}\beginargs{#2}\separgs{#3}\allargs}
\mathcommand\ifthenelseindexpp
[4]{\ifthenelseindexsymbol{#1}\beginargs{#2}\separgs{#3}\separgs{#4}\allargs}
\mathcommand\eqpp[2]{\eqsymbol\beginargs{#1}\separgs{#2}\allargs}
\mathcommand\leqpp[2]{\leqsymbol\beginargs{#1}\separgs{#2}\allargs}
\mathcommand\lespp[2]{\lessymbol\beginargs{#1}\separgs{#2}\allargs}
\mathcommand\lexlespp[2]{\lexlessymbol\beginargs{#1}\separgs{#2}\allargs}
\mathcommand\lexlimlespp[3]
               {\lexlimlessymbol\beginargs{#1}\separgs{#2}\separgs{#3}\allargs}
\mathcommand\lexpp[3]{\lexsymbol\beginargs{#1}\separgs{#2}\separgs{#3}\allargs}
\mathcommand\ackpp[2]{\acksymbol\beginargs{#1}\separgs{#2}\allargs}
\mathcommand\switchpp[1]{\switchsymbol\beginargs{#1}\allargs}
\mathcommand\swatchpp[1]{\swatchsymbol\beginargs{#1}\allargs}
\mathcommand\whilepp[2]{\whilesymbol\beginargs{#1}\separgs{#2}\allargs}
\mathcommand\nullpp[1]{\nullsymbol\beginargs{#1}\allargs}
\mathcommand\nullppiterated[2]{\nullsymbol^{#1}\beginargs{#2}\allargs}
\mathcommand\hdpp[1]{\hdsymbol\beginargs{#1}\allargs}
\mathcommand\hdppiterated[2]{\hdsymbol^{#1}\beginargs{#2}\allargs}
\mathcommand\carpp[1]{\carsymbol\beginargs{#1}\allargs}
\mathcommand\cdrpp[1]{\cdrsymbol\beginargs{#1}\allargs}
\mathcommand\tlpp[1]{\tlsymbol\beginargs{#1}\allargs}
\mathcommand\tlppiterated[2]{\tlsymbol^{#1}\beginargs{#2}\allargs}
\mathcommand\inpp[2]{\insymbol\beginargs{#1}\separgs{#2}\allargs}
\mathcommand\inppiterated[3]{\insymbol^{#1}\beginargs{#2}\separgs{#3}\allargs}
\mathcommand\applypp[2]{\applysymbol\beginargs{#1}\separgs{#2}\allargs}
\mathcommand\applyppiterated
[3]{\applysymbol^{#1}\beginargs{#2}\separgs{#3}\allargs}
\mathcommand\termpp[2]{\termsymbol\beginargs{#1}\separgs{#2}\allargs}
\mathcommand\setpp[1]{\set\beginargs{#1}\allargs}
\mathcommand\russellpp[1]{\russellsymbol\beginargs{#1}\allargs}

\mathcommand\Tpp[6]{\turingmachinesymbol\beginargs{#1}\separgs{#2}\separgs
{#3}\separgs{#4}\separgs{#5}\separgs{#6}\allargs}
\mathcommand\Tppseven[7]{\turingmachinesymbol\beginargs{#1}\separgs{#2}\separgs
{#3}\separgs{#4}\separgs{#5}\separgs{#6}\separgs{#7}\allargs}
\mathcommand\foreverppp[6]{\ident{foreverp}\beginargs{#1}\separgs{#2}\separgs
{#3}\separgs{#4}\separgs{#5}\separgs{#6}\allargs}
\mathcommand\terminatesppp[6]{\terminatespsymbol\beginargs{#1}\separgs
{#2}\separgs{#3}\separgs{#4}\separgs{#5}\separgs{#6}\allargs}
\mathcommand\terminatespppone[1]{\terminatespsymbol \beginargs{#1}\allargs}
\mathcommand\statepp
[3]{\statesymbol\beginargs{#1}\separgs{#2}\separgs{#3}\allargs}
\mathcommand\tightstatepp
[3]{\statesymbol\beginargs{#1}\tightsepargs{#2}\tightsepargs{#3}\allargs}
\mathcommand\cmdpp
[3]{\cmdsymbol  \beginargs{#1}\separgs{#2}\separgs{#3}\allargs}
\mathcommand\tightcmdpp
[3]{\cmdsymbol  \beginargs{#1}\tightsepargs{#2}\tightsepargs{#3}\allargs}
\mathcommand\stoppp           {\ident{stop}}
\mathcommand\leftpp           {\ident{left}}
\mathcommand\rightpp          {\ident{right}}
\mathcommand\nthpp         [2]{\nthsymbol  \beginargs{#1}\separgs{#2}\allargs}
\mathcommand\pppp          [1]{\ppsymbol\beginargs{#1}            \allargs}
\mathcommand\qppp          [2]{\qpsymbol\beginargs{#1}\separgs{#2}\allargs}
\mathcommand\Eppp          [1]{\Epsymbol\beginargs{#1}            \allargs}
\mathcommand\Epppzwei      [2]{\Epsymbol\beginargs{#1}\separgs{#2}\allargs}
\mathcommand\Pppp          [1]{\Ppsymbol\beginargs{#1}            \allargs}
\mathcommand\Ppppeinsindex [2]{\Ppsymbol_{#1}\beginargs{#2}\allargs}
\mathcommand\Ppppdrei      
[3]{\Ppsymbol\beginargs{#1}\separgs{#2}\separgs{#3}\allargs}
\mathcommand\Ppppvier
[4]{\Ppsymbol\beginargs{#1}\separgs{#2}\separgs{#3}\separgs{#4}\allargs}
\mathcommand\Qppp          [2]{\Qpsymbol\beginargs{#1}\separgs{#2}\allargs}
\mathcommand\Qpppeins      [1]{\Qpsymbol\beginargs{#1}\allargs}
\mathcommand\Qpppeinsindex [2]{\Qpsymbol_{#1}\beginargs{#2}\allargs}
\mathcommand\Qpppdrei      
[3]{\Qpsymbol\beginargs{#1}\separgs{#2}\separgs{#3}\allargs}
\mathcommand\Fatherpp      [2]{\Fathersymbol\beginargs{#1}\separgs{#2}\allargs}
\mathcommand\Marriespp     [2]{\Marriessymbol\beginargs{#1}\separgs{#2}\allargs}
\mathcommand\Lovespp       [2]{\Lovessymbol\beginargs{#1}\separgs{#2}\allargs}
\mathcommand\StolenBypp    [2]
{\StolenBysymbol\beginargs{#1}\separgs{#2}\allargs}
\mathcommand\Humanpp       [1]{\Humansymbol\beginargs{#1}\allargs}
\mathcommand\Evenpp        [1]{\Evensymbol\beginargs{#1}\allargs}
\mathcommand\Evenppi       [2]{\Evensymbol^{#1}\beginargs{#2}\allargs}
\mathcommand\Oddpp         [1]{\Oddsymbol\beginargs{#1}\allargs}
\mathcommand\Primepp       [1]{\Primesymbol\beginargs{#1}\allargs}
\mathcommand\EveryPairpp  [2]{\EveryPairsymbol\beginargs{#1}\separgs
{#2}\allargs}
\mathcommand\mindexppeins  [2]{\mindexsymbol{#1}\beginargs{#2}\allargs}
\mathcommand\Givepp        [3]{\Givesymbol
\beginargs{#1}\separgs{#2}\separgs{#3}\allargs}
\mathcommand\mindexppzwei  [3]{\mindexsymbol
{#1}\beginargs{#2}\separgs{#3}\allargs}
\mathcommand\mindexppdrei  [4]{\mindexsymbol
{#1}\beginargs{#2}\separgs{#3}\separgs{#4}\allargs}

\mathcommand\nonnegppp     [1]{\nonnegpsymbol\beginargs{#1}\allargs}

\mathcommand\anonymouscsymbol{c}
\mathcommand\anonymouscindexsymbol[1]{\anonymouscsymbol_{#1}}
\mathcommand\anonymousfsymbol{f}
\mathcommand\anonymouscpp
[2]{\anonymouscsymbol\beginargs{#1}\separgs\ldots\separgs{#2}\allargs}
\mathcommand\anonymouscindexpp
[3]{\anonymouscindexsymbol{#1}\beginargs{#2}\separgs\ldots\separgs{#3}\allargs}
\mathcommand\anonymousfpp
[2]{\anonymousfsymbol\beginargs{#1}\separgs\ldots\separgs{#2}\allargs}
\mathcommand\coerceindexpp[3]{[#3]_{#1}^{#2}}

\mathcommand\Elephantppp    [1]{\Elephantpsymbol\beginargs{#1}\allargs}
\mathcommand\Flowerppp      [1]{\Flowerpsymbol  \beginargs{#1}\allargs}
\mathcommand\Bicycleppp     [1]{\Bicyclepsymbol \beginargs{#1}\allargs}
\mathcommand\Germanppp      [1]{\Germanpsymbol  \beginargs{#1}\allargs}
\mathcommand\Hugeppp        [1]{\Hugepsymbol    \beginargs{#1}\allargs}
\mathcommand\Animalppp      [1]{\Animalpsymbol  \beginargs{#1}\allargs}
\mathcommand\Maleppp        [1]{\Malepsymbol    \beginargs{#1}\allargs}
\mathcommand\Boyppp         [1]{\Boypsymbol     \beginargs{#1}\allargs}
\mathcommand\Girlppp        [1]{\Girlpsymbol    \beginargs{#1}\allargs}
\mathcommand\Femaleppp      [1]{\Femalepsymbol  \beginargs{#1}\allargs}
\mathcommand\Roundppp       [1]{\Roundpsymbol   \beginargs{#1}\allargs}
\mathcommand\Bishoppp       [1]{\Bishopsymbol   \beginargs{#1}\allargs}
\mathcommand\Quadrangularppp[1]{\Quadrangularpsymbol  \beginargs{#1}\allargs}
\mathcommand\Kissedppp[2]{\Kissedpsymbol\beginargs{#1}\separgs{#2}\allargs}
\mathcommand\Metppp[2]   {\Metpsymbol   \beginargs{#1}\separgs{#2}\allargs}

\newcommand\bound     {{\rm bound}}
\newcommand\free      {{\rm free}}

\mathcommand\Vtripleindex[3]{\V\!_{{#1},\,{#2},\,{#3}}}
\mathcommand\Vdoubleindex[2]{\V\!_{{#1},\,{#2}}}
\mathcommand\Vsingleindex[1]{\V\!_{{#1}}}

\mathcommand\Erel[1]{\Gammaoffont\!_{#1}}
\mathcommand\Urel[1]{\Deltaoffont_{#1}}

\mathcommand\theRprimefromstrongtoweak{
  \inparenthesesinlinetight{
     \domres\id{\Vwall\cup\Vsome\setminus\RAN\varsigma}
     \nottight{\nottight\uplus}
     \reverserelation\varsigma
  }
  \nottight{\circ}
  \ranres
    {\transclosureinline R}
    {\Vwall\cup\Vsome\setminus\RAN\varsigma}
  \nottight{\nottight{\nottight{\uplus}}}
  \Vsome\tighttimes\Vsall
}

\mathcommand\deltaminus{\delta^-}
\mathcommand\deltaplus{\delta^+}
\mathcommand\deltaplusplus{\delta^{+^+}}
\mathcommand\deltastar{\delta^*}
\mathcommand\deltastarstar{\delta^{*^*}}

\mathcommand\Vall     {\Vsingleindex\indexdelta         }
\mathcommand\Vwall    {\Vsingleindex\indexdeltaminu     }
\mathcommand\Vsall    {\Vsingleindex\indexdeltaplus     }
\mathcommand\Vgsome   {\Vsingleindex\indexgammaplus     }
\mathcommand\Vsome    {\Vsingleindex\indexgamma         }
\mathcommand\Vfree    {\Vsingleindex\indexfree          }
\mathcommand\Vbound   {\Vsingleindex\indexbound         }
\mathcommand\Vsomesall{\Vsingleindex\indexgammadeltaplus}

\mathapplycommand\VARall      {\VARsingleindex\indexdelta         }
\mathapplycommand\VARwall     {\VARsingleindex\indexdeltaminu     }
\mathapplycommand\VARsall     {\VARsingleindex\indexdeltaplus     }
\mathapplycommand\VARgsome    {\VARsingleindex\indexgammaplus     }
\mathapplycommand\VARsome     {\VARsingleindex\indexgamma         }
\mathapplycommand\VARfree     {\VARsingleindex\indexfree          }
\mathapplycommand\VARbound    {\VARsingleindex\indexbound         }
\mathapplycommand\VARsomesall {\VARsingleindex\indexgammadeltaplus}
\mathcommand\displayVARsall[1]{\VARsingleindex\indexdeltaplus
\!\!\!\:\left(\begin{array}{@{}c@{}}#1\end{array}\right)}

\mathcommand\rigidvari     [2]{#1_{#2}^\indexgammadeltaplus}
\mathcommand\existsvari    [2]{#1_{#2}^\indexgamma    }
\mathcommand\forallvari    [2]{#1_{#2}^\indexdelta    }
\mathcommand\freevari      [2]{#1_{#2}^\indexfree     }
\mathcommand\wforallvari   [2]{#1_{#2}^\indexdeltaminu}
\mathcommand\sforallvari   [2]{#1_{#2}^\indexdeltaplus}
\mathcommand\gexistsvari   [2]{#1_{#2}^\indexgammaplus}
\mathcommand\boundvari     [2]{#1_{#2}}
\mathcommand\vari          [2]{#1_{#2}}
\mathcommand\wforallvarilow[2]{#1_{#2}^
{\raisebox{-.82ex}{\math\indexdeltaminu}}}

\newcommand\indexhelper[1]{{\scriptscriptstyle#1\:\!\!}}
\newcommand\indexdeltaplus
{\indexhelper{\delta^{\raisebox{-.17ex}{\fvesf\hskip-0.14em +}}}}
\newcommand\indexdeltaminu
{\indexhelper{\delta^{\mbox{\fvesf\hskip-0.14em\rule[.2ex]{.7em}{.15ex}}}}}
\newcommand\indexgammaplus
{\indexhelper{\gamma^{\mbox{\fvesf\hskip-0.14em +}}}}
\newcommand\indexgammadeltaplus
{\indexhelper{\gamma\delta^{\raisebox{-.17ex}{\fvesf\hskip-0.14em +}}}}

\newcommand\indexdelta{\indexhelper\delta}
\newcommand\indexgamma{\indexhelper\gamma}
\newcommand\indexfree
{{\scriptscriptstyle\free}}
\newcommand\indexbound
{{\scriptscriptstyle\bound}}

\newcommand\Wellfsymb{\ident{Wellf}}
\mathapplycommand\Wellfpp{\Wellfsymb}
\mathcommand\beginargs{(}
\mathcommand\allargs  {)}
\mathcommand\separgs  {,\,}
\mathcommand\tightsepargs{,}

\mathcommand\minusppnoparentheses  [2]{{#1}\,\minussymbol\,{#2}}
\mathcommand\tightminusppnoparentheses  [2]{{#1}\minussymbol{#2}}
\mathcommand\divideppnoparentheses [2]{{#1}\,\dividesymbol\,{#2}}
\mathcommand\plusppnoparentheses   [2]{{#1}\,\plussymbol \,{#2}}
\mathcommand\plusppnoparenthesesi  [3]{{#2}\,\plussymbol^{#1}\,{#3}}
\mathcommand\tightplusppnoparentheses   [2]{{#1}\plussymbol{#2}}
\mathcommand\timesppnoparentheses  [2]{{#1}\,\timessymbol\,{#2}}
\mathcommand\undppnoparentheses    [2]{{#1}\und            {#2}}
\mathcommand\oderppnoparentheses   [2]{{#1}\oder           {#2}}
\mathcommand\impliesppnoparentheses[2]{{#1}\implies        {#2}}
\mathcommand\leqinfixppnoparentheses[2]{{#1}\,\tight\leq\,{#2}}
\mathcommand\geqinfixppnoparentheses[2]{{#1}\,\tight\geq\,{#2}}
\mathcommand\dividepp [2]{(\divideppnoparentheses {#1}{#2})}
\mathcommand\minuspp  [2]{(\minusppnoparentheses  {#1}{#2})}
\mathcommand\pluspp   [2]{(\plusppnoparentheses   {#1}{#2})}
\mathcommand\timespp  [2]{(\timesppnoparentheses  {#1}{#2})}
\mathcommand\undpp    [2]{(\undppnoparentheses    {#1}{#2})}
\mathcommand\oderpp   [2]{(\oderppnoparentheses   {#1}{#2})}
\mathcommand\impliespp[2]{(\impliesppnoparentheses{#1}{#2})}

\mathcommand\notconflu{\mathchoice
             {{\hskip1.5pt\nmid\hskip-4.697545pt\downarrow}}%   display-style
             {{\hskip1.5pt\nmid\hskip-4.65pt\downarrow}}   %   text-style
             {{\hskip1pt\nmid\hskip-3.494pt\downarrow\hskip1pt}}  
             {{\hskip1pt\nmid\hskip-3.01pt\downarrow\hskip0.5pt}}   
}
\mathcommand\redpara{\mathchoice
           {{\redsimple\hskip-16pt  \shortparallel}\hskip8.5pt}%display-style
           {{\redsimple\hskip-16pt  \shortparallel}\hskip8.5pt}%text-style
           {{\redsimple\hskip-8.5pt \shortparallel}\hskip6pt}%scriptstyle
           {{\redsimple\hskip-7.5pt \shortparallel}\hskip5pt}%scriptscriptstyle
}
\mathcommand\antiredpara{\mathchoice
           {{\antired\hskip-14.6pt  \shortparallel}\hskip7pt}%display-style
           {{\antired\hskip-14.6pt  \shortparallel}\hskip7pt}%text-style
           {{\antired\hskip-8.pt \shortparallel}\hskip5pt}%scriptstyle
           {{\antired\hskip-7.pt \shortparallel}\hskip5pt}%scriptscriptstyle
}
\mathcommand\revpara{\mathchoice
           {{\redsimple\hskip-16pt  \infty}\hskip4.8pt}%display-style
           {{\redsimple\hskip-16pt  \infty}\hskip4.8pt}%text-style
           {{\redsimple\hskip-11.5pt\infty}\hskip4pt}%scriptstyle
           {{\redsimple\hskip-9.9pt \infty}\hskip3pt}%scriptscriptstyle
}
\mathcommand\antirevpara{\mathchoice
           {{\antired\hskip-15.4pt\infty}\hskip4pt}%display-style
           {{\antired\hskip-15.4pt\infty}\hskip4pt}%text-style
           {{\antired\hskip-10.8pt\infty}\hskip3pt}%scriptstyle
           {{\antired\hskip-9.5pt \infty}\hskip3pt}%scriptscriptstyle
}
\mathcommand\simpara{\mathchoice%this needs more fixing
           {{\redsimple\hskip-13pt  \circ}\hskip7pt}%display-style
           {{\redsimple\hskip-13pt  \circ}\hskip7pt}%text-style
           {{\redsimple\hskip-11.5pt\circ}\hskip4pt}%scriptstyle
           {{\redsimple\hskip-9.9pt \circ}\hskip3pt}%scriptscriptstyle
}

\newcommand\Proofof{Proof of}
\renewenvironment{proof}[1]{\yestop\begin{sloppypar}\noindent
{\bf\Proofof\ {#1}}\\}{\end{sloppypar}}
\renewenvironment{proofqed}[1]
{\begin{sloppypar}\def\fooqed{#1}\noindent{\bf\Proofof\ \fooqed}}
{\QEDbf\fooqed\end{sloppypar}}
\renewenvironment{proofparsep}[1]{\parindent=0pt\begin
{sloppypar}\noindent{\bf\Proofof\ {#1}}\nopagebreak\par}
{\end{sloppypar}}
\renewenvironment{proofparsepqed}[1]{\parindent=0pt\begin
{sloppypar}\def\fooqed{#1}\noindent{\bf\Proofof\ \fooqed}\nopagebreak\par}
{\nopagebreak\QEDbf\fooqed\end{sloppypar}}
\renewenvironment{proofshort}[1]{\begin{sloppypar}\noindent
{\bf\Proofof\ {#1}:}}{\end{sloppypar}}

\newcommand\germantextonquantifiereliminationwithepsilon{%
Unser zweiter vorbereitender Schritt besteht in der \germanausschaltung\ der
\germanallundseinszeichen.
Wie im vorigen Abschnitt gezeigt wurde,
k\oe nnen wir die \germananwendung\ der \germangrundformelnacommab\ \hskip.1em
und der \germanschemataalphacommabeta\ \hskip.1em 
de\es\ \germanpraedikatenkalkul\es\ mit Hilfe der \math\varepsilon-Formel und der
expliziten Definitionen \inpit{\varepsilon_1}, \inpit{\varepsilon_2} \hskip.2em
entbehrlich machen\footnotemark[1]\hspace*{-.15em}. \hskip.3em
F\ue hren wir diese \germanausschaltung\ der \germangrundformeln\
und \germanschemata\ f\ue r die \germanquantoren\ 
an der zu betrachtenden \germanableitung\ der Formel
\nlbmath{\mathfrak E} au\es\ und ersetzen wir hernach 
jeden \germanausdruck\
\nlbmath{\inpit{\mathfrak v}\,\app{\mathfrak A}{\mathfrak v}} \hskip.2em
durch \hskip.1em\maths{\displayapptight{\mathfrak A}
{\varepsilon_{\mathfrak v}\,\overline{\,\app{\mathfrak A}{\mathfrak v}}\,}},
\hskip.3em
jeden \germanausdruck\
\nlbmath{\inpit{E\,\mathfrak v}\,\app{\mathfrak A}{\mathfrak v}} \hskip.2em
durch \hskip.1em\maths{\displayapptight{\mathfrak A}
{\varepsilon_{\mathfrak v}\,\app{\mathfrak A}{\mathfrak v}}},
\hskip.4em
so gehen die au\es\ \inpit{\varepsilon_1}, \inpit{\varepsilon_2} \hskip.2em
durch \germaneinsetzung\ gewonnenen Formeln in solche \ue ber, \hskip.2em
die durch \germaneinsetzung\ 
au\es\ der Formel \nlbmath{A\hskip.12em\tightequivalent A} \hskip.2em entstehen.
\hskip.3em
Die \germanquantoren\ werden durch diese\es\ \germanverfahren\ g\ae nz\-lich
au\esi geschaltet, 
so da\sz\ {\em nunmehr gebundene Variablen au\esi schlie\sz lich in 
\germanverbindung\ mit dem \math\varepsilon-Symbol au\fti reten,
und der \germanbeweiszusammenhang\ nur durch Wieder\-holungen,
\germaneinsetzungen, \germanumbenennung\ gebundener Variablen und 
\germanschlussschemataoldspelling\ stattfindet.}}
\newcommand\englishtextonquantifiereliminationwithepsilon{%
Our second preparatory step consists in the \englishausschaltung\
of the \englishallzeichenundseinszeichenplural. \hskip.25em
As shown in the previous section, \hskip.1em
we can dispense with the \englishanwendung\ of \hskip.15em
\englishGrundformelnacommab\ 
\hskip.1em
and \englishSchemataalphacommabeta\ \hskip.1em 
of the \englishpraedikatenkalkul\
if we use the \math\varepsilon-formula and 
the explicit definitions \inpit{\varepsilon_1}, \inpit{\varepsilon_2}. \hskip.3em
If we \nolinebreak apply this \englishausschaltung\ of \englishgrundformeln\
und \englishschemata\ for the \englishquantoren\
to the formula \nlbmath{\mathfrak E} under consideration,
and afterwards \englishersetzen\ every \englishausdruckformel\
\nlbmath{\inpit{\mathfrak v}\,\app{\mathfrak A}{\mathfrak v}} \hskip.2em
\prepofersetzbar\ \hskip.1em\maths{\displayapptight{\mathfrak A}
{\varepsilon_{\mathfrak v}\,\overline{\,\app{\mathfrak A}{\mathfrak v}}\,}},
\hskip.3em
every \englishausdruckformel\
\nlbmath{\inpit{E\,\mathfrak v}\,\app{\mathfrak A}{\mathfrak v}} \hskip.2em
\prepofersetzbar\ \hskip.1em\maths{\displayapptight{\mathfrak A}
{\varepsilon_{\mathfrak v}\,\app{\mathfrak A}{\mathfrak v}}},
\hskip.4em
then the \formulae\ obtained from 
\inpit{\varepsilon_1}, \inpit{\varepsilon_2} \hskip.2em
by \englisheinsetzung\ are \englishuebergehensichwandelnppp\ into \formulae\ 
obtained by \englisheinsetzung\
from the formula \nlbmaths{A\hskip.12em\tightequivalent A}. \hskip.5em
By this \englishverfahrenprocedure, \hskip.1em
the \englishquantoren\ are completely eliminated, 
so that {\em bound variables may occur only in 
\englishverbindung\ with the \math\varepsilon-symbol,
and the \englishbeweiszusammenhangpluralisch\ may consist only of
repetitions, \englisheinsetzungen, 
\englishumbenennungdergebundenenvariablenwithof, and 
\englishschlussschemataohnemodusponens.}}

\newcommand\englishtextelevenone
{\englishtexteleventwo{Not only the notorious golden mountain is of gold, 
 but also\\the}.}
\newcommand\englishtexteleventwo[1]
{#1 round quadrangle is just as certainly round as it is quadrangular}

\newcommand\englishtextonehundredandthirteenquotation
{our translation}

\newcommand\englishtextninehundred
{Thus, in mathematics, we have no reasons to assume any meaning of
 `existence' that would be fundamentally different from that of 
 `the validity of axiomatic relations'.}

\newcommand\englishabhandlung             
{paper}

\newcommand\germanableitung               {Ab\-lei\-tung}

\newcommand\englishabstrahierenvon        
                                          {leave out of account}

\newcommand\englishaeuszere
{syntactic}

\newcommand\englishallgemeingueltigkeit 
{\index{validity!universal}universal validity}
\newcommand\englishallgemeingueltig     
{\index{validity!universal}universally valid}

\newcommand\germanallundseinszeichen
                                {\index{Allzeichen}All- und \germanseinszeichen}

\newcommand\englishallzeichenundseinszeichen
{universal and existential quantifier \englishzeichen}

\newcommand\englishallzeichenundseinszeichenplural
{\englishallzeichenundseinszeichen s}

\newcommand\englishalternativedichotomy   
{alternative}
\newcommand\englisheinealternativedichotomy
{an alternative}
\newcommand\englishalternativendichotomies
{alternatives}

\newcommand\englishangebbar               
                                          {explicitly specifiable}

\newcommand\englishangeben
                                          {explicitly specify}

\newcommand\germananwendung              {Anwendung}
\newcommand\englishanwendung             {application}

\newcommand\englishanzahlenlehre          
{theory of cardinal numbers}
\newcommand\englishauffassenpointofview   {take the point of view}              
\newcommand\englishETWASalsETWASauffassen[2]
{\englishauffassenpointofview\ that #1 is #2}

\newcommand\englishaufloesung             {\index{resolution}resolution}
\newcommand\germanausdruck                {Au\esi druck}

\newcommand\englishausdruckterm           
{expression}

\newcommand\englishausdruckformel         
{expression}

\newcommand\englishausdrueckeformeln      {\englishausdruckformel s}

\newcommand\englishausgezeichnetcanonical
{full}
\newcommand\englishAusgezeichnetcanonical
{Full}

\newcommand\propositionalcalculusindex    {\index{propostional calculus}}

\newcommand\englishAussagenKalkul         {\propositionalcalculusindex 
                                           Propositional \englishKalkul}

\newcommand\englishaussagenverbindung     
{propositional \englishverbindung}
\newcommand\englishaussagenverbindungen   
{propositional \englishverbindungen}

\newcommand\englishaussagenverknuepfung   
{propositional \englishverknuepfung}
\newcommand\englishaussagenverknuepfungen 
{propositional \englishverknuepfungen}

\newcommand\germanausschaltung            {Au\esi schaltung}
\newcommand\englishausschaltung           {elimination}

\newcommand\englishBereiche               {Domains}

\newcommand\englishbeschaffenheit
                                          {character}

\newcommand\englishbestandteil            
{part}

\newcommand\englishbeweisformalnoindex    {proof}

\newcommand\formalproofindexentry
                            {\index{proof!(formal, also called ``derivation'')}}

\newcommand\englishbeweisformal
                             {\formalproofindexentry\englishbeweisformalnoindex}

\newcommand\englishbildungsprozesse       
{formation processes}

\newcommand\englishbildungsregeln         
{formation rules}
\newcommand\englishdarstellendekonjunktion
{\index{Konjunktion!darstellende}%
 \index{conjunction, representing}%
 representing conjunction}

\newcommand\englishderjenigewelcher       {that one, which \ldots}

\newcommand\englishdual                   {\index{duality}dual}

\newcommand\englishsichdualgegenueberstehend
{\englishdual\ to each other}

\newcommand\englishdurchfuehrungimplementation
{implementation}

\newcommand\englishdurchgehen
                                          {go through}

\newcommand\englisheindeutigbestimmt
{\index{uniqueness}unique}

\newcommand\germaneinsetzung              {Ein\-setzung}
\newcommand\germaneinsetzungen            {Ein\-setzun\-gen}

\newcommand\englisheinsetzung             {substitution}

\newcommand\englisheinsetzungen           {substitutions}

\newcommand\englishentscheidungsproblemnoindex{decision \englishproblem}
\newcommand\englishentscheidungsproblem   
                   {\index{decision problem}\englishentscheidungsproblemnoindex}
\newcommand\englishEntscheidungsProblem   
                              {\index{decision problem}Decision \englishProblem}
\newcommand\englishentscheidungsprobleme  
{\englishentscheidungsproblem s}

\newcommand\englisherfahrungskomplexe     
{experience-complexes}

\newcommand\englishersetzen               {replace}

\newcommand\prepofersetzbar               {with} 

\newcommand\englishexistentialformeln     
                              {\index{existential formula}existential \formulae}

\newcommand\englishexistenzsatz            
{existence sentence}

\newcommand\englishfestumgrenzt
{well-determined and fixed}

\newcommand\englishfinit                  
                                          {finitist}
\newcommand\englishFinit                  
                                          {Finitist}

\newcommand\englishformelvariable         {formula variable}

\newcommand\englisheineemphformelvariable {a {\em\englishformelvariable}}
\newcommand\englishformelvariablen        {\englishformelvariable s}

\newcommand\englishformelsprache          
{formula language}

\newcommand\Formelaindex       {\index{Formula (a@Formula\,(a)}}

\newcommand\Formelalabelnoindex{(a)}
\newcommand\Formelalabel       {\Formelaindex\Formelalabelnoindex}
\newcommand\Formelbindex       {\index{Formula (b@Formula\,(b)}}

\newcommand\Formelblabelnoindex{(b)}
\newcommand\Formelblabel       {\Formelbindex\Formelblabelnoindex}

\mathcommand\theFormela{(x)\,A(x)\nottight\implies A(a)}

\newcommand\germangrundformelnacommab 
                               {\germangrundformeln~\Formelalabel,~\Formelblabel}
\newcommand\englishGrundformelnacommab{\Formulae~\Formelalabel,~\Formelblabel}

\mathcommand\theFormelinpiteins{a=b\nottight{\nottight{\implies}}
\inparenthesestight{a=c\nottight{\nottight{\implies}}b=c}}

\mathcommand\theFormelinpitzwei{a=b\nottight{\nottight{\implies}}b=a}

\mathcommand\theFormelinpitdrei{a=b\nottight{\nottight{\implies}}
\inparenthesestight{b=c\nottight{\nottight{\implies}}a=c}}

\newcommand\englishganzerationalefunktion 
{integer rational function}

\newcommand\englishgegenstand             
{thing}

\newcommand\englishgenetisch              
{genetic}

\newcommand\englishGeschaeftsfuehrendeHerausgeber
{Editors-in-Chief}

\mathcommand\theformelJeins{a=a}

\mathcommand\theformelJzwei{a=b\nottight{\nottight\implies}\inparenthesestight
{A(a)\nottight\implies A(b)}}

\newcommand\englishgleichheitsbeziehung
{\index{equality!relation}equality relation}

\newcommand\germangrundformel             {Grund\-formel}
\newcommand\germangrundformeln            {\germangrundformel n}
\newcommand\englishgrundformel            {\index{formula!basic}basic formula}

\newcommand\englishgrundformeln           {\englishgrundformel s}

\newcommand\englishmithilfevonexplicitly  {with the help of}

\newcommand\englishhinterglied            
{consequent}

\newcommand\englishIndividuum             {Individual}
\newcommand\englishIndividuen             {\englishIndividuum s}

\newcommand\englishindividuensymbol
                                    {\index{individual symbol}individual symbol}
\newcommand\englisheinindividuensymbol    
                                          {an \englishindividuensymbol} 
 
\newcommand\englishindividuensymbole      
                                          {\englishindividuensymbol s}
\newcommand\englishpraedikatenundindividuensymbole
                                       {predicate and \englishindividuensymbole}
\newcommand\englishpraedikatenoderindividuensymbole
                                       {predicate or \englishindividuensymbole}
\newcommand\englishwederpraedikatennochindividuensymbole
                               {neither predicate nor \englishindividuensymbole}

\newcommand\englishindividuenvariable  
                                {\index{individual variable}individual variable}
\newcommand\englisheineindividuenvariable {an \englishindividuenvariable} 
\newcommand\englishEineindividuenvariable {An \englishindividuenvariable} 
\newcommand\englishindividuenvariablen    {\englishindividuenvariable s}

\newcommand\englishjedebeliebige
                                          {any}

\newcommand\englishkalkul                 {cal\-cu\-lus}
\newcommand\englishKalkul                 {Cal\-cu\-lus}

\newcommand\englishkettenschlussindex     {\index{chain inference}}

\newcommand\englishkettenschluesse          
{\englishkettenschlussindex chain inferences}

\newcommand\englishleitgedanke
                                          {main idea}

\newcommand\englishmuendungsstelle
{leaf}
\newcommand\englishmuendungsstellen
                                          {leaves}

\newcommand\englishnachpruefen           
{check}

\newcommand\englishNamenVerzeichnis      {Index of Persons}

\newcommand\englishnormaldisjunktion
{\index{normal disjunction}normal disjunction}

\newcommand\englishnormierung            
{standardization}

\newcommand\englishnumerieren            
{number}

\newcommand\germanpraedikatenkalkul       {Pr\ae\-dikaten\-kalkul}

\newcommand\englishpraedikatenkalkul      {predicate \englishkalkul}
\newcommand\englishPraedikatenKalkul      {Predicate \englishKalkul}

\newcommand\englishEinstelligerpraedikatenkalkulohnepraedikatenkalkul
                                     {\index{predicate calculus!monadic}Monadic}

\newcommand\englishEinstelligerPraedikatenKalkul
{\englishEinstelligerpraedikatenkalkulohnepraedikatenkalkul\
                                                      \englishPraedikatenKalkul}

\newcommand\englishproblem                {problem}
\newcommand\englishProblem                {Problem}

\newcommand\germanquantor                 {Quantor}
\newcommand\germanquantoren               {\germanquantor en}

\newcommand\englishquantoren              {quantifiers}

\newcommand\Regeldeltaprimelabel
                      {\index{Rule delta'@Rule\,\inpit{\delta'}}\inpit{\delta'}}

\newcommand\englishRegeldeltaprime        {Rule\,\Regeldeltaprimelabel}

\newcommand\englishSachVerzeichnis        {Subject Index}

\newcommand\englishsatzverbindung         
{\index{sentential combination}sentential \englishverbindung}
\newcommand\englishsatzverbindungen       
{\index{sentential combination}sentential \englishverbindungen}

\newcommand\englishschemata               {schemata}
\newcommand\germanschemata                {Schemata}
\newcommand\englishSchemata               {Schemata}

\newcommand\Schemaalphaindex {\index{Schema alpha@Schema\,\inpit\alpha}}

\newcommand\Schemaalphalabelnoindex{\inpit\alpha}
\newcommand\Schemaalphalabel       {\Schemaalphaindex\Schemaalphalabelnoindex}

\newcommand\Schemabetaindex  {\index{Schema beta@Schema\,\inpit\beta}}

\newcommand\Schemabetalabelnoindex{\inpit\beta}
\newcommand\Schemabetalabel        {\Schemabetaindex\Schemabetalabelnoindex}

\newcommand\germanschemataalphacommabeta   
                          {\germanschemata~\Schemaalphalabel,~\Schemabetalabel}
\newcommand\englishSchemataalphacommabeta   
                          {\englishSchemata~\Schemaalphalabel,~\Schemabetalabel}

\newcommand\englishschluss                {inference}

\newcommand\indexenglishschlussschema{\index
  {inference schema!\edcomment{of modus ponens}}}%

\newcommand\germanschlussschemaoldspelling{Schlu\sz\-schema}
\newcommand\germanschlussschemataoldspelling{Schlu\sz\-schemata}

\newcommand\englishschlussschemaohnemodusponens
                             {\indexenglishschlussschema\englishschluss\ schema}

\newcommand\englishschlussschemataohnemodusponens{\englishschluss\ schemata}

\newcommand\germanseinszeichen            
{\index{Seinszeichen@Sein\esi zeichen}Sein\esi zeichen}
\newcommand\englishseinszeichen           
{existential quantifier \englishzeichen}

\newcommand\englishsinnlichewahrnehmung   
{sense perception}

\newcommand\englishsprachgebrauch         
{usage of language}
\newcommand\englishgewoehnlichersprachgebrauch
{common usage of language}

\newcommand\englishtatsaechlichkeit       
{actuality}
\newcommand\englishueberfuehrbar          
{\index{convertibility}convertible}

\newcommand\englishueberfuehrbarkeit      
{\index{convertibility}convertibility}

\newcommand\englishuebergehensichwandeln  {turn}

\newcommand\englishuebergehensichwandelnppp{\englishuebergehensichwandeln ed}

\newcommand\englishuebersichtlich         {easily accessible}

\newcommand\germanumbenennung             {Um\-benennung}
\newcommand\englishumbenennungboundvariables
                               {\index{renaming (of) bound variables}renaming}

\newcommand\englishumbenennungdergebundenenvariablenwithof
                         {\englishumbenennungboundvariables\ of bound variables}

\newcommand\germanverbindung              {Ver\-bindung}
\newcommand\englishverbindung             {combination}
\newcommand\englishverbindungen           {com\-bina\-tions}

\newcommand\germanverfahren               {Ver\-fahren}

\newcommand\englishverfahrenprocedure     {procedure}

\newcommand\englishverteilungvonwahrheitswerten{distribution of truth values}

\newcommand\englishprepverteilungvonwahrheitswertenauf{on}

\newcommand\englishsichverifizieren    
{prove true}

\newcommand\englishverknuepfung           
{connective}
\newcommand\englishverknuepfungen         {\englishverknuepfung s}
\newcommand\englishverknuepfungprep       
{of}

\newcommand\englishverschaerfen           
{sharpen}

\newcommand\englishschaerfer  
{sharper}

\newcommand\englishverschaerft            
{sharp\-ened}

\newcommand\englishverschaerfung          
{sharpen\-ing}

\newcommand\englishVollstaendigkeit       {\index{completeness}Completeness}

\newcommand\englishvordergliednoindex
{antecedent}

\newcommand\englishwertevorrat
{range of values}

\newcommand\englishwertsystemdervariablen 
{\index{valuation}valuation}

\newcommand\germanwertverteilungdervariablen
{Wertverteilung der Variablen}
\newcommand\englishwertverteilungdervariablen
{\index{\englishverteilungvonwahrheitswerten}%
 \index{\germanwertverteilungdervariablen}%
 \englishverteilungvonwahrheitswerten}
\newcommand\englishprepwertverteilungauf
                                   {\englishprepverteilungvonwahrheitswertenauf}

\newcommand\englishwertverteilungdervariablenexplizit
{\index{\englishverteilungvonwahrheitswerten}%
 \index{\germanwertverteilungdervariablen}%
 \englishverteilungvonwahrheitswerten\
 \englishprepwertverteilungauf\ the variables}

\newcommand\englishWiderspruchsfreiheit   {Consistency}

\newcommand\englishproblemderwiderspruchsfreiheitindex
                                                 {\index{consistency!problem of}}

\newcommand\englishProblemderWiderspruchsfreiheitnoindex
                                        {Problem of \englishWiderspruchsfreiheit}
\newcommand\englishProblemderWiderspruchsfreiheit
{\englishproblemderwiderspruchsfreiheitindex
                                   \englishProblemderWiderspruchsfreiheitnoindex}

\newcommand\englishZahlenTheorienoindex   {Number Theory}  

\newcommand\englishZahlenTheorie    
                              {\index{number theory}\englishZahlenTheorienoindex}

\newcommand\englishzahligidentisch[1]     
{\math{#1}-identical}
     
\newcommand\englishzahligidentischeins[1] 
{\englishzahligidentisch 1}
\newcommand\englishzahligidentischzwei    
{\englishzahligidentisch 2}
\newcommand\englishzahligidentischdrei    
{\englishzahligidentisch 3}
\newcommand\englishzahligidentischvier    
{\englishzahligidentisch 4}

\newcommand\englishzeichen                {symbol}

\newcommand\germanbeweiszusammenhang{Bewei\esi zusammen\-hang}
\newcommand\englishbeweiszusammenhangpluralisch{\index{connection!interconnection of the proof}interconnections of the \englishbeweisformal}

\newcommand\formulae                      {formulas}
\newcommand\Formulae                      {Formulas}

\newcommand\hbsectionhelper[1]{{\large\par\noindent\LINEnomath{{\bf #1}}\par}}
\newcommand\hbsubsectionhelper[1]{{\par\noindent\LINEnomath{{\bf #1}}\par}}
\newcommand\hbsection[2]{\hbsectionhelper{\S\,\,#1.~~~#2}}
\newcommand\hbsubsection[2]{\hbsubsectionhelper{#1.~~~#2}}
\newcommand\thehbsection[2]{\hbsection{#1}{\csname hbsection#1#2\endcsname}}
\newcommand\thehbsubsection
[3]{\hbsubsection{#2}{\csname hbsubsection#1s#2#3\endcsname}}%
\newcommand\sethbsection[4]{%
\expandafter\newcommand\csname hbsection#1#2\endcsname{#3}%
\expandafter\newcommand\csname tochbsection#1#2\endcsname{#4}%
}
\newcommand\sethbsubsection[5]{%
\expandafter\newcommand\csname hbsubsection#1s#2#3\endcsname{#4}%
\expandafter\newcommand\csname tochbsubsection#1s#2#3\endcsname{#5}%
}
\newcommand\tochbsection[3]{\contentsline
 {section}{\numberline{\S\,\,#1.}{\csname tochbsection#1#2\endcsname}}{#3}{}}
\newcommand\tochbsectionnonumber[3]{\contentsline
 {section}{\csname tochbsection#1#2\endcsname}{#3}{}}
\newcommand\tochbsubsection[3] % Only for Vol. I
{\contentsline{subsection}{\numberline{(#1)}{#2}}{#3}{}}
\newcommand\tochbIIsubsection[4]
{\contentsline
 {subsection}{\numberline{#2.}{\csname tochbsubsection#1s#2#3\endcsname}}{#4}{}}

\sethbsection{1}{I}
{The \label
{problemderwiderspruchsfreiheit1}\englishProblemderWiderspruchsfreiheit\ 
in Axiomatics \nlnomath
~~~as a Logical \englishEntscheidungsProblem}
{The \englishProblemderWiderspruchsfreiheit\ in Axiomatics
as a \\ Logical \englishEntscheidungsProblem}

\sethbsection{2}{I}
{Elementary \englishZahlenTheorie. 
 \ --- \ 
 \englishFinit\
 Inference
 \nlnomath
 ~~~and its Limits}
{Elementary \englishZahlenTheorie. 
 \ --- \ 
 \englishFinit\
 Inference
 and its Limits \ }

\sethbsection{3}{I}
{Formalization of Logical Inference I: 
 \nlnomath
 ~~~The \englishAussagenKalkul\edfootnotemark{45I1}}
{Formalization of Logical Inference I: \ 
 The \englishAussagenKalkul}

\sethbsection{4}{I}
{Formalization of Logical Inference II: 
 \nlnomath
 ~~~The \englishPraedikatenKalkul\edfootnotemark{86I1}}
{Formalization of Logical Inference II: \
 The \englishPraedikatenKalkul}

\sethbsection{5}{I}
{Adding the\edfootnotemark{163I4}  Identity. \  
 \englishVollstaendigkeit\ of the
 \nlnomath
 ~~~\englishEinstelligerPraedikatenKalkul}
{Adding the Identity.\\
 \englishVollstaendigkeit\  of the \englishEinstelligerPraedikatenKalkul}

\sethbsection{6}{I}
{\englishWiderspruchsfreiheit\ 
of  
Infinite \englishBereiche\ of \englishIndividuen.
 \nlnomath
 ~~~Beginnings of \englishZahlenTheorie.}
{\englishWiderspruchsfreiheit\
of  
Infinite \englishBereiche\ of \englishIndividuen.
\\Beginnings of \englishZahlenTheorie}

\sethbsection{7}{I}
{The Recursive Definitions}
{The Recursive Definitions}

\sethbsection{8}{I}
{The Notion ``\englishderjenigewelcher'' and its Eliminability}
{The Notion ``\englishderjenigewelcher'' and its Eliminability}

\sethbsection{9}{I}
{\englishNamenVerzeichnis}
{\englishNamenVerzeichnis}

\sethbsection{10}{I}
{\englishSachVerzeichnis}
{\englishSachVerzeichnis}

\sethbsection{1}{II}
{The Method of Eliminating the Bound Variable
 \nlnomath 
 \englishmithilfevonexplicitly\
 \hilbert's \math\varepsilon-Symbol.}
{The Method of Eliminating the Bound Variables\\\englishmithilfevonexplicitly\
 \hilbert's \math\varepsilon-Symbol}

\sethbsubsection{1}{1}{II}
{The process of the symbolic 
 \label{resolution page}%
 \englishaufloesung\ of 
 \label{existential formula page}%
 \englishexistentialformeln.}
{The process of the symbolic \englishaufloesung\ of \englishexistentialformeln}

\sethbsubsection{1}{2}{II}
{\hilbertsepsilon-symbol and the \math\varepsilon-formula.}
{\hilbertsepsilon-symbol and the \math\varepsilon-formula}

\renewcommand\namefont{\sc}
\newcommand\thispaper{this paper}
\begin{document}
\makecover
% Let your text start here, possibly changing the following a lot.
\maketitle
\begin{abstract}% MAKE SURE: No space between \begin{abstract} and abstract.
We investigate the elimination of quantifiers in first-order formulas via
\hilbert's epsilon-operator (or -binder), 
following \bernays' explicit 
definitions of the existential and the universal quantifier symbol
by means of epsilon-terms.
This elimination has its first explicit occurrence in the proof
of the first epsilon-theorem in \hilbertbernays\ in\,1939.
We think that there is a lacuna in this proof \wrt\ this elimination,
related to the erroneous assumption 
that explicit definitions always terminate.
\mbox{Surprisingly,} to \nolinebreak the best of our knowledge, 
nobody ever published a 
confluence or termination proof for this elimination procedure.
Even myths on non-confluence and the openness of the termination problem
are circulating.
We show confluence and termination of this elimination procedure by means of
a direct, straightforward, and easily verifiable proof,
based on a theorem on how to obtain termination from weak normalization.%
\end{abstract}

\vfill\pagebreak

\section{Introduction}
\halftop\subsection{The Explicit Historical Source of the Problem}
With ``\hilbertbernays'' we will designate the 
``bible of proof theory\closequotecommasmallextraspace
\ie\ the two-volume monograph {\em\grundlagendermathematik}\/
({\em Foundations of Mathematics}\/) \hskip.1em
in its two editions \makeaciteoftwo{grundlagen-first-edition-volume-one}
{grundlagen-first-edition-volume-two} and \makeaciteoftwo
{grundlagen-second-edition-volume-one}
{grundlagen-second-edition-volume-two}.

On \p 19\f\ of \cite{grundlagen-first-edition-volume-two}, \hskip.2em
\aswellas\ on \p\,20 of the second edition 
\cite{grundlagen-second-edition-volume-two}, \hskip.2em
we read:\notop\halftop
\begin{quote}\sloppy\renewcommand\tightequivalent
{\tight\sim}
``\germantextonquantifiereliminationwithepsilon''
\par\yestop\noindent
``\englishtextonquantifiereliminationwithepsilon''
\end{quote}
Note that the ``\hspace*{-.2em}\math A'' 
is not a 
meta-variable here (such as ``\math{\mathfrak A}'' is a meta-variable 
for a formula,
and ``\math{\mathfrak v}'' 
for a {\em bound \englishindividuenvariable}), \hskip.25em
but a concrete object-level \englishformelvariable. \hskip.4em
In \nolinebreak a proof step called 
{\em\englisheinsetzung}\/
either such
\englisheineemphformelvariable\ (which is always free) 
or a {\em free \englishindividuenvariable}\/
is replaced everywhere
in a formula with an arbitrary formula or term, respectively. \hskip.3em
Furthermore, 
note that ``\germanschlussschemaoldspelling''\,%
(``\englishschlussschemaohnemodusponens'') \hskip.05em
is nothing but a short name for the inference schema of {\it modus ponens.} 
\pagebreak

\begin{sloppypar}
Moreover, 
note that
\litnoteref 1 actually occurs only in the second edition
and reads
``\footnotemark[1]\Vgl\,S.15.'' (``\footnotemark[1]\Cfnlb\,\p 15.''). \hskip.4em
Neither on \litspageref{15}
---~nor anywhere else in the volumes~---
can \nolinebreak we \nolinebreak find any further information, \hskip.07em
however, \hskip.07em
regarding the following immediate \mbox{questions:}\begin{itemize}\item
In which order are the final replacements of the two explicitly mentioned forms 
of \englishausdrueckeformeln\ to be applied 
in the elimination of quantifiers?\item
Or are such eliminations independent of the order of 
the replacements in the sense that they always yield unique 
normal \nolinebreak forms?\end{itemize}%
What we can actually find on \litspageref{15} \hskip.1em
are the mentioned
``explicit definitions \inpit{\varepsilon_1}, 
\inpit{\varepsilon_2}\closequotecommaextraspace
which describe the rewrite relation of these replacements. \hskip.4em
In the more modern notation we prefer for \thispaper, \hskip.1em
these explicit definitions read:%
%\pagebreak
\par\yestop\noindent\LINEmath{\exists\boundvari x{}\stopq 
A\nottight{\nottight\equivalent}
A\{\boundvari x{}\mapsto\varepsilon\boundvari x{}\stopq A\}}
{}\mbox{~~}\inpit{\varepsilon_1}
\par\yestop\noindent\LINEmath{\forall\boundvari x{}\stopq 
A\nottight{\nottight\equivalent}
A\{\boundvari x{}\mapsto\varepsilon\boundvari x{}\stopq\neg A\}}
{}\inpit{\varepsilon_2}
\par\yestop\noindent Note that \boundvari x{} 
is a meta-variable for {\em\englishindividuenvariablen}
(in the original: a concrete object-level, bound \englishindividuenvariable), 
\hskip.2em
and \math A is a meta-variable for formulas
(in \nolinebreak the original: a \nolinebreak concrete object-level, \singulary\
\englishformelvariable). \hskip.3em
The original version of \inpit{\varepsilon_1} literally reads:
\bigmaths{(E x)\,\app A x\sim\displayapptight A{\varepsilon_x\,\app A x}}.%
\end{sloppypar}

Note that the formulas considered here and in what follows
are always \firstorder\ formulas,
extended with \math\varepsilon-terms and possibly also with 
free (\secondorder)
{\em\englishformelvariablen.} \hskip.4em
For our considerations in \thispaper,
it does not matter whether we include such \englishformelvariablen\ into 
our \firstorder\ formulas or not. \hskip.4em

\subsection{Subject Matter}
What we will study in \thispaper\ is the question
how the elimination of \firstorder\ quantifiers 
via their explicit definitions can take place.

Here we should recall that, \hskip.1em
in {\em explicit definitions}
(contrary to recursive definitions), \hskip.2em
the \nolinebreak 
symbol to be defined (here: \math\exists\ or \math\forall), \hskip.1em 
occurring 
on the left-hand side 
of an equation 
(the\,{\it definiendum}\/) \hskip.1em
must not re-occur 
in the term on the right-hand side ({\it definiens}\/). \hskip.4em

In this standard terminology, \hskip.2em
\inpit{\varepsilon_1} and \inpit{\varepsilon_2}
classify as explicit definitions, \hskip.2em
because \maths\exists{} \nolinebreak and \nlbmaths\forall{} 
do not occur on the right-hand sides
---~at least not explicitly. 
% \hskip.3em 
% (Of~course, equivalence is not identity of formulas, but this does not matter
% in our formula language here, 
% as it does not differentiate between
% formulas and propositions,
% \cf\ \eg\ \cite{Slater_2006_FregesHiddenAssumption}.)

\begin{sloppypar}
It is commonplace knowledge that
(contrary to recursive or implicit definitions) \hskip.1em 
explicit definitions are  
analytic (\ie\ not synthetic) \hskip.1em 
in the sense that they cannot contribute anything essential to our
knowledge base
\mbox{---~simply} because any notion introduced by an explicit definition
can be eliminated from any language (at least in principle) \hskip.1em
after replacing all
{\it definienda}\/ with their respective {\it definientia.}\end{sloppypar}

\pagebreak

For \firstorder\ terms the eliminability is indeed trivial,
even for non-right-linear equations such as 
\\[-1.7ex]\noindent\LINEmaths{\russellpp x=\mbppp x x},
\\[+.7ex]\noindent where the number of occurrences of defined symbols in 
\nlbmath x \hskip.1em is doubled when rewriting with this equation; \hskip.4em
\ie, \hskip.2em
if \app n t denotes the number of explicitly 
defined symbols in the term \nlbmaths t, \hskip.2em
then \bigmaths{\app n{\russellpp t}= \app n t+1},
whereas \bigmaths{\app n{\mbppp t t}\geq 2*\app n t}.

The termination of a stepwise elimination by applying 
one equation after the other
\mbox{---~until} no defined symbols remain~---
does not crucially depend on whether we rewrite the defined symbols 
in \nlbmath t \hskip.1em
before we apply the equation for the defined term \russellpp t or after.
\hskip.2em
Indeed, 
the difference this alternative can make is only a duplication 
of the rewrite steps required for the normalization of \nlbmath t. 
%\pagebreak

\begin{sloppypar}
This argumentation, \hskip.1em 
however, \hskip.1em
does not straightforwardly apply to our 
definitions \inpit{\varepsilon_1}, \nolinebreak\inpit{\varepsilon_2}. \hskip.6em
Indeed, 
the instance of the first occurrence of the meta-variable \nlbmath A 
on the right-hand side is modified by a substitution that may 
introduce an arbitrarily large number of copies of the instance of \nlbmaths A.
\end{sloppypar}

We will show in \thispaper, however, that rewriting of an arbitrary formula
\nlbmath F
with \inpit{\varepsilon_1}, \nolinebreak\inpit{\varepsilon_2} \hskip.15em
is always confluent and
terminating. \hskip.2em
This means that, 
no matter in which order we eliminate the quantifiers,
a resulting quantifier-free formula will always be obtained, 
and that this formula is a unique normal form for \nlbmath F.

\subsection{A Lacuna in \hilbertbernaysplain?}\label{section A Lacuna in}
The fact that this rewriting is innermost terminating 
has been well known before, \hskip.3em
but none of the experts on \hilbertsepsilon\ 
we consulted knew about 
the strong \mbox{termination} 
(\ie\nolinebreak\ \mbox{termination} independent of any rewriting strategy), 
\hskip.3em
and one of them even claimed that the rewriting would not be confluent.

As the proofs of the 
\math\varepsilon-theorems of \cite{grundlagen-first-edition-volume-two}
show, \hskip.1em
\bernaysname\ \bernayslifetime\ was well aware of the influence of 
strategies on elimination procedures. \hskip.3em
The mathematical technology of the\,1930s, however, makes it most unlikely
that he could easily show the strong termination ---
let alone consider it to be trivial in the context of a textbook 
(such as \hilbertbernays). \hskip.2em

Moreover,
the actual formula language of \hilbertbernays\
strongly suggests an outermost strategy: \hskip.2em
A non-outermost rewriting typically requires the instantiation of 
\nlbmath A to formulas containing variables 
that are bound by the outer quantifiers and epsilons. \hskip.3em
Such an instantiation is not permitted in \hilbertbernays, 
however,
because these additional variables must come from a set of variables
different from the free \englishindividuenvariablen,
which are called {\em bound}\/ \englishindividuenvariablen\ and 
which are not permitted to occur free 
in a substitution for \nlbmaths A. \hskip.4em
Thus, \hskip.1em
for an innermost rewriting 
in the \englishpraedikatenkalkul\ of \hilbertbernays, \hskip.1em
we have to resort to 
multiple tacit applications of \englishRegeldeltaprime\
for a complete reconstruction of the whole 
outer part of the formula in each innermost rewrite step; \hskip.3em
for \englishRegeldeltaprime\ see \eg\ \litspageref{109} in \makeaciteoftwo
{grundlagen-second-edition-volume-one}
{grundlagen-german-english-edition-volume-one-two}.

All in all, \hskip.1em
the fact that neither the innermost rewriting strategy 
nor \englishRegeldeltaprime\ 
is mentioned in this context in
\cite{grundlagen-first-edition-volume-two} \hskip.1em
makes it most likely
that \bernays\ just relied here on his learning
that explicit definitions always admit an elimination,
which is actually not the case in general for 
higher-order definitions.
\vfill\pagebreak
%%%%%%%%%%%%%%%%%%%%%%%%%%%%%%%%%%%%%%%%%%%%%%%%%%%%%%%%%%%%%%%%%%%%%%%%%%%%%%%%
\subsection{Alternative Proofs by Applying Theories of First- or Higher-Order Rewriting?}
\halftop\noindent
In \thispaper, we will approach our results directly,
without applying the theory of first- or higher-order rewrite systems. \hskip.2em
Other options for obtaining the crucial termination result 
could be:\begin{enumerate}\item
To map the \firstorder\ terms with quantifiers and epsilons
to quantifier- and epsilon-free \firstorder\ terms,
to find a \firstorder\ term rewriting system that admits
the transitive reduction of the images of any original reduction,
and to prove the termination of the \firstorder\ term rewriting system,
using the powerful theorems and methods to establish
termination of \firstorder\ term rewriting systems
(or even some of the software systems that may show \firstorder\ 
termination automatically, \cfnlb\ \eg\ \cite{middeldorp_goodstein}).
\item
To apply some results on termination 
of higher-order rewriting systems.
\item
To map the \firstorder\ terms with quantifiers and epsilons to 
\church's simply-typed \math\lambda-calculus 
(which is known to be terminating),
such that the images of each original reduction  
admit the transitive reduction in simply-typed \math\lambda-calculus.
\end{enumerate}
Let us look at \secondorder\ formulations of 
\inpit{\varepsilon_1}, \hskip.2em
partly because the original formulation of \hilbertsepsilon\ as found
in \cite{ackermann-1925} and \makeaciteoftwo{unendliche}{grundlagenvortrag}
\hskip.1em is already a \secondorder\ one without binders, \hskip.1em
and partly to develop options 2 and 3 a bit further.

If we use \math i to designate the sort (basic type) of individuals and \math o
to designate the sort of formulas
(as standard in \church's simply-typed \math\lambda-calculus), \hskip.3em
then the \nlbmath\varepsilon\ gets the typing of 
\bigmaths{\FUNDEF\varepsilon{\inpit{\FUNSET i o}}i}, and for a
\secondorder\ variable \FUNDEF A i o \hskip.2em
and the existential operator 
\FUNDEF\Sigmaoffont{\inpit{\FUNSET i o}}o, \hskip.4em
we get
\\\noindent\LINEmaths{\Sigmaoffont A=\app A{\varepsilon A}},
\\\noindent or in \math\eta-expanded form
\\\noindent\LINEmaths{\Sigmaoffont{\lambda x.\inpit{A x}}
=\app A{\varepsilon{{\lambda x.\inpit{A x}}}}}.
\par\halftop\halftop\noindent
To implement these equations according to option\,2,
we have to pick one of  
the three competing higher-order rewriting frameworks,
namely {\em combinatory reduction systems (CRSs)}
\cite{klopdiss}, \cite{klopCRS1993}, \hskip.2em
{\em higher-order rewrite systems}
\cite{nipkowLICS1991}, \cite{Raamsdonk99}, \hskip.2em
and 
{\em algebraic-functional systems} \cite{JouannaudOkadaLICS1991}. \hskip.4em
We pick the CRS framework because it is the oldest and most popular one
(also admitting extension to conditional
rewriting straightforwardly, \cfnlb\,\cite[\litnoteref 9]{wirth-jsc}).

\halftop\noindent
In CRS syntax (\cf\ \eg\ \cite[\litsectref{11}]{klopCRS1993}),
the \math\eta-expanded rule reads 
\par\halftop\noindent\LINEmaths{\Sigmaoffont{[x]\inpit{\app A x}}
=\app A{\varepsilon{{[x]\inpit{\app A x}}}}},
\par\halftop\noindent where \math x is a variable, \hskip.1em
\math A is a \singulary\ {\em meta-variable} 
(not only a top-level one, but also \wrt\ the special technical terms 
used for CRSs,
\ie\ a meta-variable for a special variable
that must not occur in the terms in the range of the rewrite relation), 
\hskip.2em
\math\Sigmaoffont\ and \math\varepsilon\ are \singulary\ function symbols
(\ie\ 1-ary constant symbols), \hskip.2em
and \math{[x]} is an abstraction operator,
binding the variable \nlbmaths x. \hskip.4em 
In this notation,
we indeed have a CRS {\em rewrite rule}\/ with the intended rewrite relation.
\hskip.05em
We can formulate \nlbmath{\inpit{\varepsilon_2}} in a similar way,
resulting in a two-rule CRS that is {\em orthogonal} 
(called ``regular'' in \cite{klopdiss}), \hskip.1em
\ie\ non-overlapping (``non-ambiguous'') and left-linear. \hskip.2em
Thus, \hskip.1em
according to \cite[\litcororef{13.6}]{klopCRS1993} 
(\cite[\littheoref{II.3.11}]{klopdiss}), \hskip.2em
the rewrite relation is confluent. \hskip.2em

As it is obvious that this rewrite relation is weakly normalizing
(as it is innermost terminating), \hskip.1em
its termination (strong normalization) 
follows from \littheoref{II.5.9.3} of \cite[\p 168]{klopdiss}, \hskip.2em
provided that we can show our rewrite relation to be non-erasing. \hskip.2em
This means that we \nolinebreak have to show that the set of free variables 
is invariant under rewrite steps. \hskip.3em
Note that the instance of 
\nlbmath A may contain free variables (such as \math y), \hskip.2em
but even if the instance of \nlbmath A \nolinebreak is, \hskip.1em say,
\bigmaths{\underline\lambda[x]\inpit{y=y}}{}
(\ie\ the quantifier is vacuous, 
binding a variable that does not occur in its scope), \hskip.2em 
it seems that the deletion of the second occurrence of \math A
in the right-hand side does not matter, \hskip.1em
because all occurrences of free 
variables are preserved by the first occurrence of \math A in the
right-hand side.

This argumentation, 
however, 
forgets that CRSs come without \math\beta-reduction. \hskip.3em
So we may need the rule
\bigmaths{\inpit{\lambda[x]\inpit{\app A x}}B
=\app A B}{} in addition, 
which would render the CRS erasing. \hskip.3em
On the other hand, 
\math{\underline\lambda} is different from \math\lambda\
(although some crucial underlining of \math\lambda\ is missing in
\cite{klopCRS1993}) \hskip.1em 
and part and parcel of the substitution framework for ``meta-variables'' in
\cite{klopCRS1993}; \hskip.3em
this means we should get along without the \math\beta-rule for \maths\lambda,
\hskip.2em
provided that we write existential quantification in our formulas as, say, 
``\math{\Sigmaoffont[x]}'' 
instead of ``\math{\Sigmaoffont\lambda x.}\closequotefullstopnospace

If the latter is indeed the case, and if our understanding of
\cite{klopdiss} is the right one, \hskip.2em
then confluence and termination
can be established by applying the theory of CRSs.

As the contacted experts on higher-order rewriting did not want to help 
settling these questions
(and no answer was found in \cite{Raamsdonk01},
\cite{raamsdonktermination} either), \hskip.2em
and as the effort to familiarize oneself (again) with the 
most fascinating and outstanding
work documented in the \PhDthesis\ \cite{klopdiss} is considerable
and disproportionate for our subject matter,
we will present here a straightforward and efficiently verifiable proof
of termination and confluence of the reduction relation 
defined directly on \firstorder\ terms with quantifiers and epsilons.

To implement option\,3, \hskip.3em
however, \hskip.1em
we could to take \math\varepsilon\ as a constant with the above
typing 
and the mapping to \church's simply-typed \math\lambda-calculus
could replace the previous constant \nlbmath\Sigmaoffont\ 
with the \math\lambda-term
\bigmaths{\lambda A\stopq\inpit{\app A{\varepsilon A}}}{}
of the same type as \math\Sigmaoffont\ before. \hskip.3em
Then reduction by the first of the above equations could be done 
by a first \math\beta-reduction, 
and a second \mbox{\math\beta-reduction}
on the \math\lambda-term \math A
could be used to reduce \app A{\varepsilon A}, \hskip.2em
such that an original reduction step with \nlbmath{\inpit{\varepsilon_1}}
results in two \math\beta-reduction steps after the mapping to 
simply-typed \math\lambda-calculus. \hskip.3em
Although this proof plan is most promising,
it is not \englishuebersichtlich\ in the sense that a mathematician could 
verify it without a careful formalization of lots of technical 
and syntactic details.
Moreover, 
as \bernays\ in the 1930s
could not have known about the termination of simply-typed
\mbox{\math\lambda-calculus}
---~first shown by \citet{tait-termination-simply-typed}~---
this is not a proof plan he could have followed
(though
he was in correspondence with \church\ and visiting the 
Institute for Advanced Study in Princeton 
during session 1935/36).

Finally, 
note that 
\ ---~~compared to options 1--3~~--- \
our direct and efficiently \mbox{verifiable}
procedure is not only more informative on the concrete structure of the
particular subject matter, but also 
the stronger, more concise, and historiographically
more relevant evidence
against myths on 
\hilbertbernays\
with regard to non-confluence and
openness of the 
termination question.%
\pagebreak

\section{Background and Tools}
\yestop\subsection{Basic Notions and Notation}
We follow standard mathematical writing style, \cf\ \cite{writing-mathematics}.

We try to be self-contained in this \thispaper. \ 
In case we should omit some required information, we refer the reader 
to the survey 
\mbox{\cite[\litsectref{I.5}]{klopdiss}}
on abstract rewrite systems.

Let `\N' denote the set of natural numbers,
and `\math<' the ordering on \nolinebreak\N\@. \
Let \maths{\posN:=\mbox{\setwith{n\tightin\N}{0\tightnotequal n}}}.

\halftop\noindent
For classes \nlbmath R, \math A, and \math B we define:\smallfootroom
\\\noindent\math{\begin{array}{@{\indent}l@{\ }l@{\ }l@{~~~~~~}l@{}}
   \DOM R
  &:=
  &\setwith{\!a}{\exists b\stopq          (a,b)\tightin R\!}
  &\mbox{{\em domain}} 
 \\\domres R A
  &:=
  &\setwith{\!(a,b)\tightin R}{a\tightin A\!}
  &\mbox{{\em(domain-) restriction to }}
   A
 \\\relapp R A
  &:=
  &\setwith{\!b}{\exists a\tightin A\stopq(a,b)\tightin R\!}
  &\mbox{{\em image of }}
   A
   \mbox{, \ \ie\ \ }
   \relapp R A=\RAN{\domres R A}
 \\\multicolumn{4}{@{}l@{}}{\mbox
 {And the dual ones:}\headroom\smallfootroom}
 \\\RAN R
  &:=
  &\setwith{\!b}{\exists a\stopq          (a,b)\tightin R\!}
  &\mbox{{\em range}}
 \\\ranres R B
  &:=
  &\setwith{\!(a,b)\tightin R}{b\tightin B\!}
  &\mbox{{\em range-restriction to \math B}}
 \\\revrelapp R B
  &:=
  &\setwith{\!a}{\exists b\tightin B\stopq(a,b)\tightin R\!}
  &\mbox{{\em reverse-image of }}
   B
   \mbox{, \ \ie\ \ }
   \revrelapp R B=\DOM{\ranres R B}
 \\\end{array}}
\par\noindent
We \nolinebreak 
use `\id' for the
identity function, and `\math\circ' for the composition of binary relations.
Functions are (right-) unique relations, \hskip.1em 
and so
the meaning of \hskip.15em
``\nolinebreak\hspace*{-.1em}\nolinebreak\math
{f\tight\circ g}\nolinebreak\hskip.15em\nolinebreak'' 
is extensionally given by
 \bigmaths{\app{\inpit{f\tight\circ g}}x=\app g{\app f x}}. \hskip.2em

% Furthermore, \hskip.1em
% we use `\math\emptyset' to denote the empty set as well as the
% empty function. \hskip.4em
Let \redsimple\ be a binary relation. \ 
\redsimple\ is a relation {\em on \math A} \udiff\ 
\mbox{\math{\DOM\redsimple
\nottight\cup\RAN\redsimple\subseteq A.}} \ 
\redsimple\ \nolinebreak is \nolinebreak{\em irreflexive}\/ \udiff 
\bigmaths{\id\cap\redsimple=\emptyset}. \
It is \math A{\em-reflexive\/} \udiff
\bigmaths{\domres\id A\subseteq\redsimple}. \
Speaking of a {\em reflexive}\/ relation
we refer to the largest \math A that is appropriate in the local context,
and referring to this \math A
we write \redindexn 0{} 
to ambiguously denote 
\math{\domres\id A}. \ 
With \math{\redindexn 1{}:=\redsimple}, and
\math{\redindexn{n+1}{}:=\redindexn n{}\tight\circ\redsimple} 
for \math{n\in\posN}, \ 
\redindexn m{} \nolinebreak 
denotes the \math m-step relation
for \nlbmath\redsimple. \ 
The {\em transitive closure}\/ of \nlbmath\redsimple\ is 
\bigmaths{\trans:=\bigcup_{n\in\posN}\redindexn n{}}. \ 
The {\em reflexive closure}\/ of \nlbmath\redsimple\ is 
\bigmath{\onlyonce:=\bigcup_{n\in\{0,1\}}\redindexn n{}.}
The {\em reflexive transitive closure}\/ of \nlbmath\redsimple\ is 
\bigmath{\refltrans:=\bigcup_{n\in\N}\redindexn n{}.}
The {\em reverse}\/ of \redsimple\ is
\math{\antired:=\setwith{\pair b a}{\pair a b\tightin\redsimple}}.

\math v \nolinebreak and \nlbmath w are called
{\em joinable \wrt\ \nlbmath\redsimple} \udiff\ \math{v\tight\downarrow w}, \ 
\ie\ \udiff\ \math{v\refltrans\circ\antirefltrans w}\@. \ \
\redsimple\ is {\em locally confluent} \udiff\
\maths{v\tight\downarrow w}{} \hskip.2em
for any \maths v, \math w 
with \bigmaths{v\antired\circ\redsimple w}; 
it is {\em confluent} \udiff\
\maths{v\tight\downarrow w}{} \hskip.2em
for any \maths v, \math w 
with \bigmaths{v\antirefltrans\circ\refltrans w}.
\math{a'} is an {\em\redsimple-normal form of \math a}
\udiff\ \bigmaths{a\refltrans a'\notin\DOM\redsimple}.

A sequence \bigmath{\inpit{s_i}_{i\in\N}}
is {\em non-terminating in \redsimple}\/
\udiff\ \bigmath{s_i\redsimple s_{i+1}} for all \math{i\in\N}.
\ \redsimple\ \nolinebreak is {\em terminating}\/
\udiff\ there are no non-terminating sequences in \nlbmath\redsimple. \ 
\mbox{A relation \math R (on \math A)} is \mbox{\em\wellfounded}\/ 
\udiff\ 
any non-empty class \math B (\math{\tightsubseteq A}) has an
\math R-minimal element, \ie\ \math
{\exists a\tightin B\stopq\neg\exists a'\tightin B\stopq a' R\,a}. \ 
Note that \wellfoundedness\ of \antired\ immediately entails termination 
of \redsimple\ (via the range of the non-terminating sequence), \hskip.2em
but the converse requires a weak form of the \axiomofchoice\
to construct the non-terminating sequence, \hskip.2em
\cf\ \eg\ \cite
[\litsectref{4.1}]{Moore_Wirth_Automation_of_Mathematical_Induction_2013}.

\begin{corollary}\label{corollary wellfounded} 
If a binary relation is \wellfounded, so is its transitive closure.
\end{corollary}%
\pagebreak
% Let \math<\ be the irreflexive ordering of the natural numbers \nlbmath\N.
% Let \redsimple\ denote a binary relation.
% \begin{definition}[{Increasing \cite[\litdefiref{I.5.16(2)}, \p 52]{klopdiss}}]
% \\\redsimple\ is {\em increasing} \udiff\
% there is a map \FUNDEF c{\FLD\redsimple}\N,
% \\such that, for any \math s, \math t with \bigmaths{s\redsimple t}, 
% we have \bigmaths{\app c s<\app c t}.\end{definition}
%%%%%%%%%%%%%%%%%%%%%%%%%%%%%%%%%%%%%%%%%%%%%%%%%%%%%%%%%%%%%%%%%%%%%%%%%%%%%%%%
%%%%%%%%%%%%%%%%%%%%%%%%%%%%%%%%%%%%%%%%%%%%%%%%%%%%%%%%%%%%%%%%%%%%%%%%%%%%%%%% 
\subsection{A  Generalized Theorem as the Main Tool}
\noindent The following \theoref{theorem a la klop} 
is a generalization of \klopname's \littheoref{I.5.18}
\cite[\p\,53]{klopdiss}, which can be obtained again 
from \theoref{theorem a la klop} by 
the specialization \nlbmaths{\redindex 0:=\emptyset}.
\begin{theorem}\label{theorem a la klop}
\\Let \redindex 0 and \redindex 1 be two binary relations. \hskip.4em
\\Set \maths{\redindex 2:=\refltransindex 0\circ\redindex 1}. \hskip.4em
\\Set \maths{\redindex 3:=\redindex 0\cup\redindex 1}. \hskip.4em
\\Let\/ \maths{a\in\DOM{\redindex 3}}. \hskip.4em
\mbox{Let\/ \math{a'} be} an \redindex 3-normal form of\/ \maths a. \hskip.4em
Set \math{A:=\relappsin{\refltransindex 3}a}. \hskip.4em
\\Set \maths{\redindex 4:=\domres{\redindex 3}A}. \hskip.4em
If\begin{enumerate}\noitem\item
\ranres{\antiredindex 0}A is \wellfounded;\noitem\item
there is an upper bound\/ \math{n\in\N} on the length of\/ 
\redindex 2-derivations starting from\/ \nlbmath a and reaching \nlbmath{a'} 
by \refltransindex 0; \hskip.5em
more formally, \hskip.2em
this means that we have \maths{m\leq n}{}
for any\/ \math{m\tightin \N} and any 
sequence
\nlbmath{b_0,\ldots,b_m} 
with\/ 
\maths{a\tightequal b_0}, \hskip.3em
\maths{\,b_i\redindex 2 b_{i+1}}{}
for each
\maths{i\in\{0,\ldots,m\tight-1\}}, \hskip.4em and\/ 
\maths{b_m\refltransindex 0 a'};\noitem\item
for all\/ \math{b_1,b_2} with
\bigmaths{b_1\antiredindex 4\circ\redindex 1 b_2}, 
we have \bigmaths{b_1\refltransindex 4\circ\antirefltransindex 4 b_2}; 
and\noitem\item
for all\/ \math{b_1,b_2} with
\bigmaths{b_1\antiredindex 4\circ\redindex 0 b_2}, 
we have \bigmathnlb{b_1\refltransindex 4\circ\antionlyonceindex 4 b_2};\noitem
\end{enumerate}
then\/ \hskip.1em 
\antiredindex 4 is \wellfounded.\end{theorem}
%%%%%%%%%%%%%%%%%%%%%%%%%%%%%%%%%%%%%%%%%%%%%%%%%%%%%%%%%%%%%%%%%%%%%%%%%%%%%%%% 
\yestop\begin{proofparsepqed}{\theoref{theorem a la klop}}
\initial{\underline{\underline{Claim\,1:}}}
For all \math{b_1,b_2} and \math{n\in\N} with
\bigmaths{b_1\antiredindex 4\circ\redindexn n 0 b_2}, 
we have \bigmathnlb{b_1\refltransindex 4\circ\antionlyonceindex 4 b_2}.
\initial{\underline{\underline{Proof of Claim\,1:}}}
By induction on \nlbmaths n. \hskip.3em
In case of \bigmaths{b_1\antiredindex 4\circ\redindexn 0 0 b_2}, 
we have \bigmaths{b_1\antiredindex 4 b_2}. In case of 
\bigmaths{b_1\antiredindex 4\circ\redindexn n 0 b_2\redindex 0 b_3}, 
by induction hypothesis we have 
\bigmathnlb{b_1\refltransindex 4 b_4\antionlyonceindex 4 b_2}{} 
for some \maths{b_4\in A}. \hskip.4em
In case of \maths{b_4\tightequal b_2}, \hskip.2em
we have \bigmaths{b_1\refltransindex 4 b_4\redindex 0 b_3},
and thus \bigmaths{b_1\refltransindex 4 b_3}.
Otherwise, we have \maths{b_4\antiredindex 4 b_2}, \hskip.2em
and thus \bigmaths{b_4\refltransindex 4 b_5\antionlyonceindex 4 b_3}{}
for some \math{b_5} by \lititemref 4, \hskip.5em
\ie\ the desired \bigmaths{b_1\refltransindex 4 b_5\antionlyonceindex 4 b_3}.
\QEDdouble{Claim\,1}
\par
Set \maths{B:=\setwith{b\tightin A}{b\refltransindex 4 a'}}. \hskip.4em
\par By \lititemref 2, we can define a function 
\FUNDEF{l}{B}{\setwith{m\tightin\N}{m\leq n}} \hskip.2em via 
\par\noindent\LINEmaths{\app l b := \max\setwith{m\in\N}
{b\redindexn m 2\circ\refltransindex 0 a'}}.
\initial{\underline{\underline{Claim\,2:}}}
For all \math{b\in B} \hskip.2em
with \bigmaths{b\refltransindex 4 b'},
we have \bigmaths{b'\in B}.
\initial{\underline{\underline{Proof of Claim\,2:}}} 
% By an induction on the number of steps, using Claim\,1.
By induction on \nlbmaths{k:=\app l b}{} in \nlbmath<. \hskip.4em
The induction hypothesis is that
for all \math{b''\in B} \hskip.2em
with \bigmaths{b''\refltransindex 4 b'''}{} and 
\bigmaths{\app l{b''}< k},
we have \bigmaths{b'''\in B}. \hskip.3em
Note that (for \math{b''\in B}) 
\bigmaths{b''\redindex 4 b'''}{} implies
\bigmaths{\app l{b'''}\leq\app l{b''}}.
Thus, by another induction on the length of derivations,
the induction conclusion follows from the induction hypothesis and 
the proposition that
for all \math{b''\in B} \hskip.2em
with \bigmaths{b''\redindex 4 b'''}{} and 
\bigmaths{\app l{b''}=k},
we have \bigmaths{b'''\in B}.
\\So let us assume \bigmaths{b\in B}{}
and \bigmaths{b\redindex 4 b'}.
Then,
using the induction hypothesis,
we have to show \bigmaths{b'\tightin B},
for which it suffices to show \bigmaths{b'\refltransindex 4 a'}.
\\By our assumption, we have \bigmaths{b\refltransindex 4 a'},
which falls into at least one of the following two cases:
\initial{\underline{\maths{b\refltransindex 0 a'}:}}
By Claim\,1:
\bigmaths{b'\refltransindex 4\circ\antionlyonceindex 4 a'}.
Because \maths{a'\tightnotin\DOM{\redindex 3}}, \hskip.2em
and \afortiori\ also \maths{a'\tightnotin\DOM{\redindex 4}}, \hskip.2em
we actually have \bigmaths{b'\refltransindex 4 a'}.

\initial
{\underline{\maths{b\refltransindex 0\hat b\redindex 1 b'''
\refltransindex 4 a'}{} for some \maths{\hat b}, \maths{b'''}:}}
Again by Claim\,1, \hskip.1em
we get \bigmaths{b'\refltransindex 4 b''''\antionlyonceindex 4\hat b}{}
for some \maths{b''''\in A}. \hskip.8em
\\In case of \bigmaths{b''''\tightequal \hat b},
we have
\bigmaths{b'\refltransindex 4 b''''\redindex 1 b'''\refltransindex 4 a'},
\ie\ the desired
\bigmaths{b'\refltransindex 4 b''''\redindex 4 b'''\refltransindex 4 a'}.
\\Otherwise we have \bigmaths{b''''\antiredindex 4\hat b}.
Thus, by \lititemref 3, \hskip.2em
there is some \math{b''} with 
\bigmaths{b''''\refltransindex 4 b''\antirefltransindex 4 b'''}.
Because of 
\bigmaths{b\refltransindex 0\hat b\redindex 1 b'''\refltransindex 4 a'}{}
we have \bigmaths{b'''\tightin B}{} and \bigmaths{\app l{b'''}<\app l b}.
Thus, by the induction hypothesis,
we get \bigmaths{b''\tightin B},
and then the desired 
\bigmaths{b'\refltransindex 4 b''''\refltransindex 4 b''\refltransindex 4 a'}.
\QEDdouble{Claim\,2}\par
\initial{\underline{\underline{Claim\,3:}}} \bigmaths{A\tightequal B}.
\initial{\underline{\underline{Proof of Claim\,3:}}} 
By \bigmaths{a\refltransindex 3 a'}, \hskip.2em 
we also have \bigmaths{a\refltransindex 4 a'}, \hskip.2em 
and so \bigmaths{a\tightin B}.
\\Thus, by Claim\,2, \hskip.1em
we get \bigmaths{\relappsin{\refltransindex 4}a\subseteq B}.
\\All in all, 
we get: 
\LINEmaths{A
=\relappsin{\refltransindex 3}a
=\relappsin{\refltransindex 4}a
\subseteq B\subseteq A}.
\QEDdouble{Claim\,3}\par
By Claim\,4, \hskip.1em
we get \bigmaths{\FUNDEF{l}{A}{\setwith{m\tightin\N}{m\leq n}}}. 
Now for every \math{b_1}, \math{b_2} \hskip.1em with 
\bigmaths{b_1\antiredindex 4 b_2},
we \nolinebreak have \bigmaths{b_1,b_2\tightin A}{}
and, moreover, 
\pair{\app l{b_1}}{b_1} is strictly smaller than \pair{\app l{b_2}}{b_2}
in the lexicographic combination of \nlbmath<\
and \ranres{\antiredindex 0}A, 
which is \wellfounded\ by \lititemref 1. \hskip.5em
Indeed, in case of \bigmaths{b_1\antiredindex 0 b_2},
we have \bigmaths{\app l{b_1}\leq\app l{b_2}}{} 
and \bigmaths{b_1\ranres{\antiredindex 0}A b_2},
and in case of \bigmaths{b_1\antiredindex 1 b_2},
we have \bigmaths{\app l{b_1}<\app l{b_2}}.
\end{proofparsepqed}%%%%%%%%%%%%%%%%%%%%%%%%%%%%%%%%%%%%%%%%%%%%%%%%%%%%%%%%%%%%
\yestop\subsection{Terms, Formulas, Substitutions, Contexts}
A straightforward intuitive understanding of terms,
formulas, substitutions, and contexts will actually suffice for most
working mathematicians to understand the remainder of \thispaper. \hskip.3em
For the others, \hskip.1em
we give an example formalization of these notions here.
\halftop\par\noindent
{\em Terms}\/ and {\em formulas}\/ are defined inductively 
\asfollows:\begin{itemize}\noitem\item
\englishEineindividuenvariable\ is a term.\noitem\item
If \hskip.1em\math A is an \math n-ary \englishformelvariable\
(\math{n\tightin\N}) \hskip.1em
and \math{t_1,\ldots,t_n} are terms,\\
then \math{A(t_1,\ldots,t_n)} is a formula.\noitem\item
If \hskip.1em\math{\rm f} is an \math n-ary constant function or predicate 
symbol (\math{n\tightin\N}) \hskip.1em
and \math{t_1,\ldots,t_n} are terms,\\
then \math{{\rm f}(t_1,\ldots,t_n)} is a term or formula, \hskip.1em
respectively.\\
In case of \maths{n\tightequal 0}, \hskip.1em
we simply write ``\math{{\rm f}\hskip.05em}'' instead of \hskip.15em
``\math{{\rm f}()}\closequotefullstopnospace\noitem\item
If \math F is a formula, \hskip.1em
then \math{\neg F} is a formula. \hskip.5em
If \math{F_1} and \math{F_2} are formulas, \hskip.2em
then\\\maths{\inpit{F_1\tightoder F_2}}, \hskip.2em
\maths{\inpit{F_1\tightund F_2}}, \hskip.2em
\maths{\inpit{F_1\tightimplies F_2}}, \hskip.2em
\ldots\ are formulas.\noitem\item
If \math x is \englisheineindividuenvariable\ and \math F is a formula,
\\then \mbox{\math{\varepsilon x\stopq F}} \hskip.2em is a term
and \mbox{\math{\exists x\stopq F}} and \mbox{\math{\forall x\stopq F}} 
are formulas.\\
In these terms and formulas, 
all occurrences of \math x are {\em bound\/}; \hskip.1em
non-bound occurrences of variables in terms and formulas 
are called {\em free,}\/
such as each occurrence of any \englishformelvariable, and also of 
any \englishindividuenvariable\
\nlbmath y that is not in the scope of a binder on \nlbmaths y, \hskip.2em
such as \hskip.1em ``\math{\varepsilon y.}'', \hskip.2em
``\math{\exists y.}'', \hskip.2em
or \hskip.1em``\math{\forall y.}''.
\end{itemize}
\par\noindent
In our definition of terms and formulas we deviate from \hilbertbernays\
in not having an extra set of \englishindividuenvariablen\ for bound occurrences,
\hskip.1em disjoint from the set to be used for free occurrences. \hskip.3em
So we have only one set of \englishindividuenvariablen, \hskip.1em
but this does not really make any difference here, \hskip.1em
in particular because we ignore the variable names in the bound occurrences
by the following stipulation:

{\em We equate formulas modulo the renaming of bound variables.}

A {\em substitution}\/ is a mapping of \englishindividuenvariablen\
to terms and of \math n-ary \englishformelvariablen\ to 
expressions of the form \bigmaths{\underline\lambda(x_1,\cdots,x_n)\stopq F},
respectively, \hskip.2em
where \math{x_1,\cdots,x_n} are mutually distinct 
\englishindividuenvariablen\ and \math F is a formula. \hskip.4em
For \maths{n\tightequal 0}, \hskip.2em
we just write ``\math{\!F\hskip.1em}'' instead of \hskip.15em
``\math{\underline\lambda()\stopq F\hskip.1em}\closequotefullstopnospace

Presupposing the above stipulation of considering formulas only up to 
renaming of bound variables,
we now define the result of an application of a substitution \nlbmath\sigma\
to terms and formulas inductively \asfollows. \hskip.3em
We use postfix notation with highest operator precedence.
% for substitution application.
\begin{itemize}\item
Let \nlbmath x be \englisheineindividuenvariable.\\
If \maths{x\tightnotin\DOM\sigma}, \hskip.2em
then \bigmaths{x\sigma=x}; \hskip.3em
otherwise \bigmaths{x\sigma=\app\sigma x},
\ie\ the value of \math x under \nlbmaths\sigma.\item
Let \hskip.1em\math A be an \math n-ary \englishformelvariable, \hskip.2em
and let \math{t_1,\ldots,t_n} be terms. \hskip.3em
If \maths{A\tightnotin\DOM\sigma}, \hskip.2em
then \bigmaths{\inpit{A(t_1,\ldots,t_n)}\sigma=
A(t_1\sigma,\ldots,t_n\sigma)}.
Otherwise \bigmath{\inpit{A(t_1,\ldots,t_n)}\sigma}{}
is the result of the \mbox{\math\beta-reduction} of 
\bigmaths{\app\sigma A(t_1\sigma,\ldots,t_n\sigma)},
\ie, \
for \bigmaths{\app\sigma A=\underline\lambda(x_1,\cdots,x_n)\stopq F},
the formula \maths{F\sigma'}, \
where \math{\sigma'} is the \mbox{substitution 
\maths{\{x_1\mapsto t_1\sigma\comma\ldots\comma x_n\mapsto t_n\sigma\}}.}\item 
If \hskip.1em\math{\rm f} is an \math n-ary constant function or predicate 
symbol and \math{t_1,\ldots,t_n} are terms,\\
then \bigmaths
{\inpit{{\rm f}(t_1,\ldots,t_n)}\sigma
={\rm f}(t_1\sigma,\ldots,t_n\sigma)}.\item
If \math F is a formula, \hskip.1em
then \bigmaths{\inpit{\neg F}\sigma=\neg F\sigma}. \hskip.5em
If \math{F_1} and \math{F_2} are formulas, \hskip.2em
then\\ 
\bigmaths{\inpit{F_1\tightoder F_2}\sigma
=\inpit{F_1\sigma\tightoder F_2\sigma}},
\bigmaths{\inpit{F_1\tightund F_2}\sigma
=\inpit{F_1\sigma\tightund F_2\sigma}}, 
\bigmaths{\inpit{F_1\tightimplies F_2}\sigma
=\inpit{F_1\sigma\tightimplies F_2\sigma}}, \ \ldots\@.\item
If \nlbmath x \nolinebreak is 
\englisheineindividuenvariable\ 
\\---~~\wrog\ neither an element of \DOM\sigma, \hskip.2em
nor occurring
(free) in \RAN\sigma~~---\\ 
and \math F is a formula,\\
then \bigmathnlb{\inpit{\varepsilon x\stopq F}\sigma=
\varepsilon x\stopq F\sigma}, 
\bigmathnlb{\inpit{\exists x\stopq F}\sigma=
\exists x\stopq F\sigma}, 
\bigmathnlb{\inpit{\forall x\stopq F}\sigma=
\forall x\stopq F\sigma}.\end{itemize}

\begin{corollary}\ 
If\/ \math X is \englisheineindividuenvariable\ or a nullary 
\englishformelvariable, and\/ \math\sigma\ is a substitution,
then for any formula or term\/ \math G whose free variables are in \nlbmath A: 
\bigmaths{G\sigma=G\inpit{\domres\sigma A}}.
\end{corollary}
By induction on the construction of \nlbmath{G_1} we easily get:
\begin{corollary}\label{corollary substitution B}
\\For any term or variable\/ \nlbmaths{G_1}, \hskip.2em
any\/ \math X and\/ \math{G_2} being either \englisheineindividuenvariable\
and a term, or a nullary \englishformelvariable\ and a formula,
and any substitution \nlbmaths\sigma\ where\/ \math{X\tightnotin\DOM\sigma} 
and\/ \math X does not occur (free)
in \maths{\RAN\sigma}: 
\bigmaths{\inpit{G_1\{X\tight\mapsto G_2\}}\sigma
=\inpit{G_1\sigma}\{X\tight\mapsto G_2\sigma\}}.\end{corollary}

\yestop\noindent
Finally, let \math{H_0,\ldots,H_n} \inpit{n\tightin\N} be mutually distinct,
nullary \englishformelvariablen,
reserved for the following definition: \hskip.3em
A {\em context}\/ written ``\math{G[\cdots]}''
(a formula or term with holes) \hskip.1em
is actually a formula or term \nlbmath G with one single (free) occurrence of 
each of the \englishformelvariablen\ \nlbmaths{H_1,\ldots,H_n}. \hskip.5em
Moreover, \hskip.3em
``\math{G[F_1,\ldots,F_n]}'' \hskip.3em
denotes 
\bigmaths{G\{H_1\tight\mapsto F_1\comma\ldots\comma H_n\tight\mapsto F_n\}},
for formulas \nlbmaths{F_1,\ldots,F_n}.

\begin{corollary}\label{corollary substitution A}
\\For any context \nlbmaths{G[\cdots]}, \hskip.2em
and any formula \nlbmaths F, \hskip.2em
and any substitution \nlbmaths\sigma:
\bigmaths{\inpit{G[F]}\sigma=G\sigma[F\sigma]}.\pagebreak\end{corollary}
%%%%%%%%%%%%%%%%%%%%%%%%%%%%%%%%%%%%%%%%%%%%%%%%%%%%%%%%%%%%%%%%%%%%%%%%%%%%%%%% 
% \noindent
% As a corollary from \theoref{theorem a la klop},
% we get our main tool:
% \begin{theorem}[{\cite[\litcororef{I.5.19(i)}, \p\,55]{klopdiss}}]
% \\If\/ \redsimple\ is locally confluent, weakly normalizing,
% and increasing,
% \\then\/ \redsimple\ is confluent and terminating.\end{theorem}
%%%%%%%%%%%%%%%%%%%%%%%%%%%%%%%%%%%%%%%%%%%%%%%%%%%%%%%%%%%%%%%%%%%%%%%%%%%%%%%% 
\section{The Concrete Rewrite Relation}
By writing ``\math{\neg^\forall}'' for ``\math{\neg}'' and 
``\math{\neg^\exists}'' for the empty string ``\,'', \hskip.4em
we can unify the two formulas \inpit{\varepsilon_1} and \inpit{\varepsilon_2} 
to the single formula 
\par\halftop\noindent\LINEmaths
{Q\boundvari x{}\stopq A\nottight{\nottight\equivalent}
A\{\boundvari x{}\mapsto\varepsilon\boundvari x{}\stopq\neg^Q A\}}
{}\inpit{\varepsilon_Q}{}
\par\halftop\noindent for \maths{Q\in\{\exists,\forall\}}, \hskip.3em
and \math x a meta-variable for an \englishindividuenvariable, \hskip.2em 
and \math A a meta-variable for a formula.

Let \redsimple\ be the rewrite relation 
resulting from rewriting with the equivalence 
\inpit{\varepsilon_Q} as a rewrite rule from left to right. \hskip.3em
Explicitly, 
this means that \math{F_1\redsimple F_2} \hskip.2em if 
there are a context \nlbmaths{G[\cdots]}, \hskip.2em
a quantifier symbol \nlbmaths Q, \hskip.2em
an \englishindividuenvariable\ \nlbmaths x, \hskip.2em
and a formula \maths A, \hskip.2em
such that \bigmaths{F_1=G[Q x\stopq A]}{}
and \bigmathnlb{F_2=G[A\{x\mapsto\varepsilon x\stopq \neg^Q A\}]}.

Let \redindex 0 and \redindex 1
be the partition of \redsimple\ for the case
of a {\em vacuous}\/ quantifier
(\ie\ for the case that \math x does not occur in the formula \nlbmath A 
in \inpit{\varepsilon_Q}), \hskip.2em
and for the case that the quantifier is not vacuous.

Let \redindex{\cal I} 
be the innermost rewrite relation given by rewriting with the 
equivalence \nlbmaths{\inpit{\varepsilon_Q}}.

Let \redparaindex{} be the version of \nlbmath\redsimple\
for the rewriting of parallel redexes. 
Explicitly,
this means that \math{F_1\redparaindex{}F_2} \hskip.2em if 
there are a context \nlbmath{G[\cdots]} with \math{n\in\N} holes,
quantifier symbols \maths{Q_1,\ldots,Q_n}, \hskip.2em
\englishindividuenvariablen\ \maths{x_1,\ldots,x_n}, \hskip.2em
and formulas \maths{A_1,\ldots,A_n}, \hskip.2em
such that
\par\noindent\LINEmaths{\begin{array}{l l l}F_1
 &=
 &G[Q_1 x_1\stopq A_1\comma\ldots\comma Q_n x_n\stopq A_n],
\\\mediumheadroom F_2
 &=
 &G[A_1\{x_1\mapsto\varepsilon x_1\stopq \neg^{Q_1}A_1\}\comma\ldots\comma 
    A_n\{x_n\mapsto\varepsilon x_n\stopq \neg^{Q_n}A_n\}].
\\\end{array}}{}

\halftop\noindent
From these definitions, we immediately get the following corollaries.
\begin{corollary}\label{corollary innermost is always}\bigmaths
{\redindex{\cal I}\subseteq\redsimple}.\end{corollary}
\begin{corollary}\label{corollary para refltrans}\bigmaths
{\redparaindex{}\subseteq\refltrans}.\end{corollary}

% Let the maps \maths{\rm q}, \math{\rm v}, \maths{\rm e}, 
% be given as the number 
% of occurrences of quantifier symbols of a formula,
% vacuous quantifier symbols,
% \math\varepsilon-symbols, respectively.
%%%%%%%%%%%%%%%%%%%%%%%%%%%%%%%%%%%%%%%%%%%%%%%%%%%%%%%%%%%%%%%%%%%%%%%%%%%%%%%%
\subsection{Local Confluence}
Note that the technical terms of the following lemma are clarified and
formalized in its proof.
\begin{lemma}\label{lemma overlapping locally confluent}
If we have a peak \bigmaths{F_1\antired F_0\redsimple F_2}{}
of local divergence
and the redex of the rewrite step to \nlbmath{F_1} is properly inside the one 
of the rewrite step to \nlbmath{F_2} (which is on top of \nlbmath{F_0}), 
\hskip.2em
then there are formulas \nlbmath{F_3,F_4} satisfying all the following items:
\begin{enumerate}\noitem\item
\maths{F_1\redsimple F_4\antired F_3\antiredparaindex{}F_2}.\item
If the initial step to the left is actually applied to a 
non-vacuous quantifier 
\\\mbox{(\ie\ if \math{F_1\antiredindex 1 F_0}),} \hskip.3em
then we have \maths{F_4\antiredindex 1 F_3\antiredparaindex 1 F_2}.\item 
If the initial step to the right is actually applied to a 
non-vacuous quantifier 
\\\mbox{(\ie\ if \math{F_0\redindex 1 F_2}),} \hskip.3em
then we have \maths{F_1\redindex 1 F_4}.\item 
If the initial step to the right is actually applied to a 
vacuous quantifier 
\\\mbox{(\ie\ if \math{F_0\redindex 0 F_2}),} \hskip.3em
then we have \maths{F_3=F_2}.\pagebreak\end{enumerate}\end{lemma}
%%%%%%%%%%%%%%%%%%%%%%%%%%%%%%%%%%%%%%%%%%%%%%%%%%%%%%%%%%%%%%%%%%%%%%%%%%%%%%%%
%%%%%%%%%%%%%%%%%%%%%%%%%%%%%%%%%%%%%%%%%%%%%%%%%%%%%%%%%%%%%%%%%%%%%%%%%%%%%%%%
\begin{proofparsepqed}{\lemmref{lemma overlapping locally confluent}}
Suppose we have a peak \bigmaths{F_1\antired F_0\redsimple F_2}{}
of local divergence
and the redex of the rewrite step to \nlbmath{F_1} is properly inside the one 
of the rewrite step to \nlbmaths{F_2},
which is on top of \nlbmath{F_0}. \hskip.4em
Then \math{F_0} has the form 
\\[-1.3ex]\noindent\LINEmaths
{Q_1\boundvari x 1\stopq G_1[Q_2\boundvari x 2\stopq G_2]}.%
\inpit{F_0}
\par\noindent We may in particular assume here 
that \boundvari x 2 is different from \boundvari x 1 and does not occur free in 
the context \nlbmaths{G_1[\cdots]}{} \hskip.1em
if we \nolinebreak consider the dots ``\math\cdots'' to be empty. \hskip.3em
Moreover we may assume that the formulas \math{F_1} and \math{F_2}
are the following:
\mathcommand\substitutionconfluenceeinsinnerterm
{G_1[G_2\{\boundvari x 2\mapsto\varepsilon\boundvari x 2\stopq\neg^{Q_2}G_2\}]}
\mathcommand\substitutionconfluenceeins
{\inbraces{\boundvari x 1\mapsto\varepsilon\boundvari x 1\stopq\neg^{Q_1}
\substitutionconfluenceeinsinnerterm}}%
\par\noindent\LINEmaths{Q_1\boundvari x 1\stopq 
G_1[G_2\{\boundvari x 2\mapsto\varepsilon\boundvari x 2\stopq\neg^{Q_2}G_2\}]}.%
\inpit{F_1}
\par\noindent\LINEmaths{
\inparentheses{G_1[Q_2\boundvari x 2\stopq G_2]}
\inbraces{\boundvari x 1\mapsto\varepsilon\boundvari x 1\stopq\neg^{Q_1}
G_1[Q_2\boundvari x 2\stopq G_2]}}.\inpit{F_2}%
\par\noindent 
If we rewrite the outermost redex in \nlbmaths{F_1}, \hskip.2em
we obtain the formula
\par\noindent\LINEmaths{
\inparentheses
{G_1[G_2\{\boundvari x 2\mapsto\varepsilon\boundvari x 2\stopq\neg^{Q_2}G_2\}]}
\sigma}{}
\par\noindent written with the help of the substitution \nlbmath\sigma\ given as
\par\noindent\mbox{~~~~~~~~~}\LINEmaths\substitutionconfluenceeins.\inpit\sigma
\par\noindent
If we propagate this substitution, 
by \cororef{corollary substitution A} \hskip.1em
we obtain a formula given by the context
\par\noindent\LINEmaths{G_1\sigma[\cdots]}{}\inpit{C}
\\\noindent where we read the dots ``\math\cdots'' as
\par\noindent\LINEmaths{
\inpit{G_2\{\boundvari x 2\mapsto\varepsilon\boundvari x 2\stopq\neg^{Q_2}G_2\}}
\sigma}.
\par\noindent 
Because \math{\boundvari x 2} 
occurs free in none of \nlbmaths{\DOM\sigma}, \hskip.2em
\nlbmath{G_1[\cdots]}, \hskip.2em
\substitutionconfluenceeinsinnerterm, \hskip.2em 
\RAN\sigma, \hskip.2em
by \cororef{corollary substitution B} \hskip.1em
we can propagate \nlbmath\sigma\ \hskip.1em further to write the inner formula as
\par\noindent\mbox{~~~}\LINEmaths{G_2\sigma\{\boundvari x 2\mapsto
\varepsilon\boundvari x 2\stopq\neg^{Q_2}G_2\sigma\}}.\inpit I
\par\noindent Putting \inpit C and \inpit I together again, \hskip.1em
we can choose formula \nlbmath{F_4} 
with the property \bigmaths{F_1\redsimple F_4}{}
\asfollows:
\\[-.9ex]\noindent\LINEmaths
{G_1\sigma\inbrackets{G_2\sigma\{\boundvari x 2\mapsto
\varepsilon\boundvari x 2\stopq\neg^{Q_2}G_2\sigma\}}}.\inpit{F_4}
%%%%%%%%%%%%%%%%%%%%%%%%%%%%%%%%%%%%%%%%%%%%%%%%%%%%%%%%%%%%%%%%%%%%%%%%%%%%%%%%
\par\noindent
If we now rewrite all occurrences of the redex 
mentioned at the end of the notation of the formula \nlbmath{F_2}
in parallel, \hskip.1em
then we obtain the formula
\par\noindent\LINEmaths{
\inparentheses{G_1[Q_2\boundvari x 2\stopq G_2]}\sigma}.
\par\noindent Before we can rewrite the remaining redex, \hskip.1em
we have to propagate \nlbmath\sigma\ \hskip.1em
to obtain a clear description of it. \hskip.2em 
By \cororef{corollary substitution A}, \hskip.2em
this results again in a context as given in \inpit{C}
above, \hskip.1em
where, \hskip.1em
however, \hskip.1em
we \nolinebreak now read the ``\math\cdots'' as
\par\noindent\LINEmaths{Q_2\boundvari x 2\stopq G_2\sigma}.
\par\noindent 
Note that, \hskip.1em
in this formula, \hskip.1em
the substitution \nlbmath\sigma\ has passed the quantifier 
``\math{Q_2\boundvari x 2.}'' soundly. \hskip.3em
Indeed, \hskip.1em
as \nolinebreak mentioned above, \hskip.2em 
\boundvari x 1 is different from \boundvari x 2, \hskip.2em and
\math{\boundvari x 2} cannot occur free in \RAN\sigma. \hskip.4em
Putting this formula and its context together again, 
we can choose as \nlbmath{F_3} with the property 
\bigmaths{F_3\antiredparaindex{}F_2}{}
\asfollows:
\\[-.9ex]\noindent\LINEmaths{G_1\sigma\inbrackets{Q_2\boundvari x 2\stopq G_2\sigma}}.%
\inpit{F_3}
\par\noindent
If we now rewrite the remaining redex, 
we \nolinebreak again obtain the formula \nlbmaths{F_4}, as was to be shown
for \lititemref{1}. 

For \lititemref{2}, \hskip.2em
it suffices to note that, \hskip.1em
if \boundvari x 2 occurs free in \nlbmaths{G_2}, \hskip.1em
then \boundvari x 2 also occurs free in \nlbmath{G_2\sigma}
because \boundvari x 1 \nolinebreak and \nlbmath{\boundvari x 2} are different.

For \lititemref{3}, \hskip.2em
it suffices to note that, \hskip.1em
if \boundvari x 1 occurs free in
\nlbmaths{G_1[Q_2\boundvari x 2\stopq G_2]}, \hskip.1em
then \boundvari x 1 also occurs free in \math
{G_1[G_2\{\boundvari x 2\mapsto\varepsilon\boundvari x 2\stopq\neg^{Q_2}G_2\}]}.

For \lititemref{4}, \hskip.2em
it suffices to note that, \hskip.1em
if \boundvari x 1 does not occur free in
\nlbmaths{G_1[Q_2\boundvari x 2\stopq G_2]}, \hskip.1em
then both \math{F_2} and \nlbmath{F_3} are 
actually \nlbmaths{G_1[Q_2\boundvari x 2\stopq G_2]}.
\end{proofparsepqed}
%%%%%%%%%%%%%%%%%%%%%%%%%%%%%%%%%%%%%%%%%%%%%%%%%%%%%%%%%%%%%%%%%%%%%%%%%%%%%%%%
\yestop\yestop\yestop\noindent As overlaps are trivial and 
as peaks of local divergence with parallel redexes are joinable in one step 
at each side trivially, \hskip.1em
we \nolinebreak get 
as a corollaries of \lemmref{lemma overlapping locally confluent}(1,4)
and \cororef{corollary para refltrans}:
\begin{corollary}\label{corollary epsilon is confluent}\ 
\redsimple\ is locally confluent.\end{corollary}%
\begin{corollary}\label{corollary epsilon is confluent vacuous}\ 
For all \maths{F_1}, \maths{F_2}{} 
with \bigmaths{F_1\antired\circ\redindex 0 F_2},
we have \bigmaths{F_1\trans\circ\antired\,F_2}.\end{corollary}
%%%%%%%%%%%%%%%%%%%%%%%%%%%%%%%%%%%%%%%%%%%%%%%%%%%%%%%%%%%%%%%%%%%%%%%%%%%%%%%%
\halftop\halftop
%%%%%%%%%%%%%%%%%%%%%%%%%%%%%%%%%%%%%%%%%%%%%%%%%%%%%%%%%%%%%%%%%%%%%%%%%%%%%%%%
\subsection{\WellFoundedness}
\halftop\noindent
As every \redindex 0-step (vacuous quantifiers)
and every \redindex{\cal I}-step (innermost quantifiers)
reduces the number occurrences of quantifiers by 
\nlbmaths 1, \hskip.2em we have:
\begin{corollary}\label{corollary 0 wellfounded}
\bigmaths{\antiredindex 0\cup\antiredindex{\cal I}}{} is \wellfounded.%
\end{corollary}

\yestop\begin{theorem}\label{theorem epsilon terminates}\ 
\antired\ is \wellfounded.\end{theorem}
\begin{proofparsepqed}{\theoref{theorem epsilon terminates}}
Assume that \math B is a non-empty class. \hskip.2em
Then there is some \math{a\in B}. \hskip.3em
We just have to find an \antired-minimal element in \nlbmaths B.

If \math a is \antired-minimal in \nlbmaths B, \hskip.2em
then we have succeeded. \hskip.2em
Thus suppose that \math a is not \antired-minimal in \nlbmaths B. \hskip.4em
Then \maths{a\in\DOM\redsimple}. \hskip.5em

Set \maths{A:=\relappsin\refltrans a}. \hskip.5em
Set \maths{\redindex 4:=\domres\redsimple A}. \hskip.5em
It \nolinebreak now suffices to show that \antiredindex 4 is \wellfounded\
(because an \antiredindex 4-minimal element of \nlbmath{A\cap B} is also
 an \antired-minimal element of \nlbmath B).

By \cororef{corollary 0 wellfounded}, \hskip.2em
\math A has an \antiredindex{\cal I}-minimal element \nlbmaths{a'}. \hskip.5em
As \math{a'\tightnotin\DOM\redsimple} by \cororef{corollary innermost is always},
\hskip.2em
\math{a'} \nolinebreak is an \redsimple-normal form of \nlbmaths a. \hskip.5em
To obtain the \wellfoundedness\ of \antiredindex 4, \hskip.2em
we are now going to apply \theoref{theorem a la klop}.\par
Set \maths{\redindex 2:=\refltransindex 0\circ\redindex 1}. \hskip.5em
Set \maths{\redindex 3:=\redindex 0\cup\redindex 1}. \hskip.5em
Then \bigmaths{\redsimple=\redindex 3}.\par
It now suffices to show \lititemfromtoref 1 4 
of \theoref{theorem a la klop}. \hskip.3em 
\litItemref 1 holds by \cororef{corollary 0 wellfounded}. \hskip.3em 
\litItemref 3 holds by \cororef{corollary epsilon is confluent}. \hskip.3em 
\litItemref 4 holds by \cororef{corollary epsilon is confluent vacuous}. 
\hskip.3em 
As the number of occurrences of the \nlbmath\varepsilon\ is 
invariant under \redindex 0 and is increased at least by \nlbmath 1 \hskip.1em
by every \redindex 1-step, it increases at least by \nlbmath 1 \hskip.1em
by every \redindex 2-step. Thus, to satisfy \lititemref 2, \hskip.1em
we can choose the upper bound \nlbmath n to be the number of 
occurrences of \nlbmath\varepsilon\ in \nlbmath{a'}
(minus the number in \nlbmath a).\end{proofparsepqed}
\vfill\pagebreak

\yestop\subsection{Confluence}
By the \newmanlemma\ (\cf\ \cite{newman} or, for a formal proof,
\cite[\litsectref{3.4}]{wirthcardinal}), \hskip.2em
we obtain from \cororef{corollary epsilon is confluent} and
\theoref{theorem epsilon terminates}:
\begin{theorem}\label{theorem epsilon confluent}\ 
\redsimple\ is confluent.\end{theorem}

\subsection{On the Length of Derivations}
By \theorefs{theorem epsilon terminates}{theorem epsilon confluent}, \hskip.2em
we now know for certain that the rewrite relation is confluent and terminating
(as its reverse is even \wellfounded), \hskip.2em
which means that we can eliminate the quantifiers
in any order
---~but this does not mean that this is efficient. \hskip.3em

Here is a serious warning to the contrary: \hskip.2em
The nesting depth of the occurrences of the \math\varepsilon-symbols
introduced by the normalization 
can be exponential in the number of quantifiers in the 
input formula, and the number of steps of an outermost
normalization is even higher and seems to be non-elementary, 
\cfnlb\ \cite[\litexamref{4.7}]{SR--2011--01},
\cite[\litexamref 8]{wirth-jal}.

As any innermost rewrite step reduces the number of quantifiers 
exactly by \nlbmaths 1, \hskip.2em
and as no rewrite step can reduce the number of quantifiers by more than 
\nlbmath 1, \hskip.2em 
we immediately get:%
\begin{theorem}\label{theorem length of derivations}\\ 
Let \math F be a formula with \math n~quantifiers. \hskip.3em
Innermost rewriting 
of \nlbmath F by \redindex{\cal I}
obtains the (unique) \redsimple-normal form \nlbmath{F'} of\/ \nlbmath F 
in exactly\/ \hskip.1em \math n~steps, \hskip.2em
which is the minimal number of steps to reach \nlbmath{F'} by \redsimple\
from \nlbmaths F.%
\end{theorem}
%%%%%%%%%%%%%%%%%%%%%%%%%%%%%%%%%%%%%%%%%%%%%%%%%%%%%%%%%%%%%%%%%%%%%%%%%%%%%
\section{Conclusion}
With \theorefs{theorem epsilon terminates}{theorem epsilon confluent}, \hskip.2em
we have shown confluence and termination of the elimination of 
quantifiers via their explicit definition via \hilbertsepsilon. \hskip.3em
This means in particular that any \firstorder\ term with
quantifiers and epsilons (and \englishformelvariablen), \hskip.2em
has a unique normal form \wrt\ this elimination of quantifiers, \hskip.1em
which has its first explicit occurrence
in \cite{grundlagen-first-edition-volume-two}, \hskip.2em
namely in the proof of the \nth 1 \math\varepsilon-theorem on \litspageref{19\f}

Moreover,
the directness, self-containedness, and easy verifiability of the proofs
should settle the questions on confluence and termination here
once and for all
\mbox{---~at} \nolinebreak least for working mathematicians. \hskip.4em
Formalists and rewriters, however, 
may see the need to develop a more formal verification of our proof
and write a short paper that our results are
all trivial in some higher-order rewriting theory.
\hskip.3em
Writing or helping to find a good textbook on higher-order rewriting,
however,
seems to be in more urgent demand.

% Also note that our new 
% \theoref{theorem a la klop} \hskip.1em
% may be a helpful tool also in other cases,
% in particular because it seems that the theory for obtaining 
% termination from weak normalization in abstract rewrite systems
% is still very poor.

Furthermore,
we hope that some philosophers will be stimulated by \thispaper\
to pick up the subject of the non-triviality
of higher-order {\em explicit definition}\/ and write or help to find a book 
on that subject.

Finally,
the starting point of our interest in the subject, \hskip.1em
namely the question whether there is a lacuna in 
\hilbertbernays\ as discussed in \sectref{section A Lacuna in}, \hskip.3em
needs further discussion by the experts on \hilbertsepsilon\
and the history of mathematical logic in the \nth{20}\,century. \hskip.4em
On \nolinebreak basis \nolinebreak of our current knowledge,
we would clearly answer this question positively.
\vfill\pagebreak

\catcode`\@=11
\renewcommand\@openbib@code{%
      \advance\leftmargin\bibindent
      \itemindent -\bibindent
      \listparindent \itemindent
      \parsep 10pt
      \itemsep .5pt
      \baselineskip 15pt
      }%
\catcode`\@=12
\bibliography{herbrandbib}
\end{document}